\documentclass[lettersize,journal]{IEEEtran}
\usepackage{amsmath,amsfonts}
\usepackage{algorithmic}
\usepackage{array}
\usepackage[caption=false,font=normalsize,labelfont=sf,textfont=sf]{subfig}
\usepackage{textcomp}
\usepackage{stfloats}
\usepackage{url}
\usepackage{verbatim}
\usepackage{graphicx}
\usepackage{balance}
\usepackage{booktabs} 
\usepackage{pgfplots}
\usepackage{soul}
\usepackage{amsmath}
\usepackage{gensymb}
\usepackage{flushend} 
\usepackage{threeparttable}
\usepackage{subcaption}
\usepackage{subfig}
\pgfplotsset{compat=1.18}
\usepackage{tikz}
\usetikzlibrary{shapes.geometric, arrows, positioning}
\usepackage{siunitx}
\usepackage[colorlinks=true, allcolors=black]{hyperref}
\usepackage{orcidlink}
\usepackage{siunitx}
\DeclareSIUnit{\Mbps}{\mega\bit\per\second}
\DeclareSIUnit{\kbps}{\kilo\bit\per\second}

\def\BibTeX{{\rm B\kern-.05em{\sc i\kern-.025em b}\kern-.08em
    T\kern-.1667em\lower.7ex\hbox{E}\kern-.125emX}}
\usepackage{balance}

\begin{document}
\title{Battery-Free and Gateway-Free Cellular IoT Water Leak Detection System}

\author{
    Roshan Nepal\orcidlink{0009-0007-2038-2420}, \IEEEmembership{Graduate Student Member, IEEE},
    Brandon Brown,
    Shishangbo Yu, \IEEEmembership{Graduate Student Member, IEEE},
    Roozbeh Abbasi,
    Norman Zhou\orcidlink{0000-0003-2886-0259},
    and George Shaker\orcidlink{0000-0002-1450-2138}, \IEEEmembership{Senior Member, IEEE}
\thanks{Manuscript received xxxx; revised xxxx; accepted xxxx. Date of publication xxxx; date of current version xxxx.}
\thanks{Roshan Nepal, Shishangbo Yu, and George Shaker are with the Wireless Sensors and Devices Lab (WSDL), Department of Electrical and Computer Engineering, University of Waterloo, Canada (email: roshan.nepal@uwaterloo.ca; s342yu@uwaterloo.ca; gshaker@uwaterloo.ca).}
\thanks{Brandon Brown, Roozbeh Abbasi, and Norman Zhou are with the Centre for Advanced Materials Joining (CAMJ), Department of Mechanical and Mechatronics Engineering, University of Waterloo, N2L 3V9 Waterloo, Canada (email: bbrown@uwaterloo.ca; rabbasi@uwaterloo.ca; nzhou@uwaterloo.ca).}% <-this % stops a space
}

\maketitle

\begin{abstract}
This paper presents a battery-free and gateway-free water leak detection system capable of direct communication over LTE-M (Cat-M1). The system operates solely on energy harvested through a hydroelectric mechanism driven by an electrochemical sensor, thereby removing the need for conventional batteries. To address the stringent startup and operational power demands of LTE-M transceivers, the architecture incorporates a compartmentalized sensing module and a dedicated power management subsystem, comprising a boost converter, supercapacitor-based energy storage, and a hysteresis-controlled load isolation circuit. This design enables autonomous, direct-to-cloud data transmission without reliance on local networking infrastructure. Experimental results demonstrate consistent LTE-M beacon transmissions triggered by water-induced energy generation, underscoring the system’s potential for sustainable, maintenance-free, and globally scalable IoT leak detection applications in smart infrastructure.
\end{abstract}


%\sethlcolor{yellow}
\section{Introduction}
\IEEEPARstart{W}{ater} leakage is a pervasive challenge with significant societal, economic, and environmental consequences across residential, commercial, and industrial settings. Undetected leaks accelerate structural deterioration, promote mold growth, disrupt industrial operations, and cause major financial losses~\cite{epa_fix_a_leak,Kadu2015}. At a global scale, approximately 126 billion m\textsuperscript{3} of treated water are lost every year through distribution system leakages, representing an economic cost of about US \$39 billion annually~\cite{hawle2024}. In many urban networks, this corresponds to nearly 30\% of the total water supplied, and in some aging infrastructures, losses can reach 50\%~\cite{veolia2024}. These inefficiencies not only burden utilities and consumers but also undermine efforts toward sustainable water management, climate resilience, and the United Nations Sustainable Development Goal (SDG) 6: Clean Water and Sanitation~\cite{un_sdg_report_2025}. Addressing leakage through reliable, early detection is therefore critical to global water conservation and infrastructure sustainability. These escalating concerns have spurred a demand for autonomous, maintenance‐free systems that detect leaks early and alert stakeholders in real time.

Recent advances in the Internet of Things (IoT) have introduced wireless leak-detection solutions that can send notifications or alarms immediately once a leak is sensed~\cite{Alshami2024,Ali2022,Bakhtawar2023,MaheshKumar2022}. In IoT deployments, continuous monitoring, early alerts, and data-driven maintenance are enabled by linking distributed sensors to cloud analytics and automated notifications\cite{Islam2022,Jan2022}. At scale, leak-monitoring networks must provide wide-area coverage, reliable event delivery, long operational lifetimes, and low total cost of ownership, while remaining simple to deploy, provision, and manage across large fleets of nodes. These needs span sectors: in residential and commercial buildings, devices should be unobtrusive and low maintenance; utilities require district- and city-scale visibility; industrial plants need high reliability and integration with existing maintenance systems; and remote or infrastructure-sparse sites need autonomous operation.

A wide range of wireless technologies has been employed for leak detection within IoT networks, each offering unique advantages and trade-offs. Bluetooth Low Energy (BLE) provides simple configuration and very low power consumption for short-range communication, but its dependence on nearby hubs or smartphones restricts large-scale deployment~\cite{Koulouras2025, Jeon2018, Sultania2023}. Zigbee and other IEEE 802.15.4-based protocols support mesh networking that can cover multi-room or building-scale areas, yet they require a central coordinator or gateway to connect to the internet~\cite{Zohourian2023, AlShuhail2022}. LoRa and LoRaWAN extend communication range to several kilometers at minimal power, but rely on specialized gateways and network servers for backhaul connectivity~\cite{Chilamkurthy2022,Nikoukar2018, Delgado2021}. Alternative strategies like Wi‐Fi backscatter and EnOcean focus on extremely minimal transmissions or mechanical harvesting but are typically limited to short distances or specialized infrastructure \cite{Toro2022, Gong2024,enocean2020whitepaper}.

Although these technologies demonstrate impressive efficiency within their respective domains, they share a common constraint: dependence on batteries for power and gateways for data relay. Batteries restrict operational lifetime and necessitate periodic replacement, which increases ownership costs and creates environmental waste~\cite{Hasan2023,Jamshed2022,Cao2024}. In large-scale deployments, maintaining thousands of sensor nodes requires frequent on-site servicing or truck rolls for battery replacement and disposal, which further raises labor, travel, and compliance costs. Likewise, gateway-dependent architectures require installation, reliable power, and network backhaul for intermediary hubs that relay data to the cloud~\cite{Choudhary2024,Jeon2018,Zachariah2015}. These gateways also require periodic firmware and security updates, credential management, and monitoring to maintain stable communication\cite{Michalski2021}. While such systems perform well in controlled or infrastructure-rich environments, they do not scale efficiently to remote or large-area deployments where accessibility and maintenance costs are high. Consequently, there remains a pressing need for a truly infrastructure-free sensing platform that eliminates both battery replacement and gateway dependence, enabling simpler, more sustainable, and large-scale leak monitoring.

To achieve energy autonomy, recent research has focused on harvesting ambient energy from diverse sources such as solar, thermal, vibrational, RF, and electrochemical~\cite{Safaei2025,Liu2021,AkinPonnle2021,Naifar2024}. Each source exhibits distinct power and availability characteristics: solar provides comparatively high power but depends on illumination; thermal harvesting requires stable temperature gradients; vibration harvesting is site specific and intermittent; RF harvesting offers wide accessibility but yields very low power density; and electrochemical or hydroelectric harvesting generates power directly from fluid contact. In the context of leak detection, hydroelectric or electrochemical harvesting is particularly advantageous because the presence of water at the leak site naturally triggers the reactions needed for energy generation~\cite{Rouhi2024,Nepal2024,nepal_Lora}. This event-driven mechanism produces energy precisely when and where it is required, enabling maintenance-free operation and eliminating the need for continuous external power. However, the wireless uplink remains a major challenge since radio transmissions demand short, high-power bursts and elevated voltages that far exceed the steady, low-level output typical of energy harvesters~\cite{Pereira2020,Georgiou2018,Bedi2018}. As a result, most battery-free implementations employ ultra-low-power protocols such as BLE or LoRa, which reduce transmission energy at the cost of limited range and continued reliance on gateways for backhaul connectivity.

\begin{table*}[t]
\centering
\caption{Comparison of Wireless Technologies for Infrastructure Monitoring}
\label{tab:tech_comparison}
\begin{tabular}{lllll}
\toprule
Protocol & Range & Infrastructure & Peak Current & Deployment Limitation \\
\midrule
BLE / Zigbee & Short ($<$100 m) & Gateway / Smartphone & Low ($\sim$10 mA) & Requires local hub \& power \\
LoRaWAN & Long (km) & Specialized Gateway & Low ($\sim$40 mA) & Gateway installation \& backhaul \\
LTE-M (Terrestrial) & Wide Area & Cellular Tower & High ($\sim$300 mA) & Coverage gaps in remote areas \\
LTE-M (Satellite) & Global & LEO Satellite & Max ($\sim$350+ mA) & Extreme power budget \\
\bottomrule
\end{tabular}
\end{table*}

\begin{table*}[t]
\centering
\caption{Comparison of Communication Constraints in Terrestrial and Satellite IoT Networks}
\label{tab:sat_comparison}
\begin{tabular}{lccc}
\toprule
\textbf{Parameter} & \textbf{Terrestrial LTE-M} & \textbf{LEO Satellite (NTN)} & \textbf{Legacy GEO Satellite} \\
\midrule
Orbit Altitude & 0 km (Ground) & 300 -- 1,200 km & $\sim$36,000 km \\
Relative Velocity & 0 km/s & $\sim$7.5 km/s & 0 km/s \\
Latency & $<$15 ms & 20 -- 40 ms & $>$600 ms \\
Doppler Shift & Negligible & High ($\pm$40 kHz) & Negligible \\
Path Loss & Low ($\sim$2 km range) & High ($\sim$600 km range) & Very High \\
Tx Power Req. & Variable (Low to Max) & Max (23 dBm) & High (Proprietary) \\
Hardware & Standard Modem & Standard Modem (FW update) & Proprietary Terminal \\
\bottomrule
\end{tabular}
\end{table*}

Moving beyond local gateways, the emergence of low-power cellular standards such as NB-IoT and LTE-M (both categorized under 5G LPWAN) offers direct-to-cloud connectivity using existing mobile networks~\cite{Moges2023,Vaezi2022}. These carrier-grade protocols operate in licensed spectrum and provide secure, SIM-based authentication with standardized provisioning, which simplifies large-scale deployment in utility, industrial, and infrastructure monitoring applications~\cite{Moges2023,Vaezi2022,Borkar2020}. Although originally optimized for terrestrial networks, a major shift occurred with 3GPP Release 17, which formalizes IoT operation over Non-Terrestrial Networks (NTN)~\cite{Saad2024, 3gpp_rel17}. Table~\ref{tab:tech_comparison} compares these cellular standards against legacy short-range protocols, illustrating the trade-off between infrastructure independence and power demand.

Historically, satellite IoT relied on Geostationary (GEO) satellites positioned at roughly 36{,}000~km altitude~\cite{Centenaro2021, Raghunandan2022}. The long propagation distance imposed severe path loss and latency, requiring high-power proprietary terminals and large antennas that were unsuitable for small embedded devices. Modern Low Earth Orbit (LEO) constellations such as SpaceX Starlink and Amazon Project Kuiper operate at significantly lower altitudes between 300 and 2,000~km~\cite{Osoro2021, Qu2017}. The reduced distance improves the link budget and lowers latency, enabling a Direct-to-Device (D2D) communication model in which standard LTE-M modems can connect to satellites that function as orbiting cellular base stations~\cite{Osoro2021}, as summarized in Table~\ref{tab:sat_comparison}. However, unlike stationary cell towers, LEO satellites move at high velocities ($\sim$7.5 km/s) relative to the ground~\cite{elikbilek2022}. This creates significant Doppler shifts and variable propagation delays that the device must actively compensate for using GNSS-based pre-correction and standardized timing advance protocols~\cite{Wang2021}. Furthermore, because the link distance is still significantly larger than terrestrial cells ($\sim$600 km vs 5 km), the user equipment (UE) is almost invariably forced to operate at its maximum transmission power class (23 dBm) to close the uplink budget. This capability extends cellular IoT coverage to remote, offshore, and infrastructure-sparse environments where terrestrial connectivity is limited or absent.

However, enabling this broad connectivity comes with a substantial energy cost. Whether communicating with a terrestrial base station or a LEO satellite, LTE-M transceivers require elevated startup currents and higher operating voltages~\cite{Yang2021, Labdaoui2023}. These requirements exceed what most energy harvesting sources can supply. The power profile is also bursty and difficult to predict, influenced by network timing, paging intervals, and varying signal conditions, which makes it particularly challenging to support without batteries or a large energy reservoir~\cite{Bhilwaria2021}.

As a result, achieving reliable, fully battery-free cellular operation remains a distinct challenge compared to lower-power alternatives. While battery-free architectures have been successfully implemented for protocols like BLE and LoRaWAN, these solutions inherently depend on local gateways for backhaul connectivity~\cite{Jeon2018, Delgado2021}. In the cellular domain, a small number of experimental systems have demonstrated battery-free NB-IoT operation under highly constrained conditions. These implementations typically rely on harvesting ambient energy from continuous external sources, such as indoor light~\cite{Sultania2023}, industrial waste heat~\cite{AragonesOrtiz2020,Aragones2022}, or triboelectric nanogenerators~\cite{Hu2021}. However, these sources are often intermittent or site-specific, and uncorrelated with the sending event itself. Furthermore, the higher power requirements of LTE-M have largely precluded its adoption in battery-free sensors. Sustaining LTE-M therefore requires careful energy buffering and power-path control so that harvested energy can be accumulated and released to meet the modem’s surge currents during the short but critical transmission window. To the best of our knowledge, a fully battery-free, hydro-powered system capable of establishing a direct LTE-M uplink without any local gateway has not previously been demonstrated.

\begin{figure*}[t]
\centering
\includegraphics[width=0.95\textwidth]{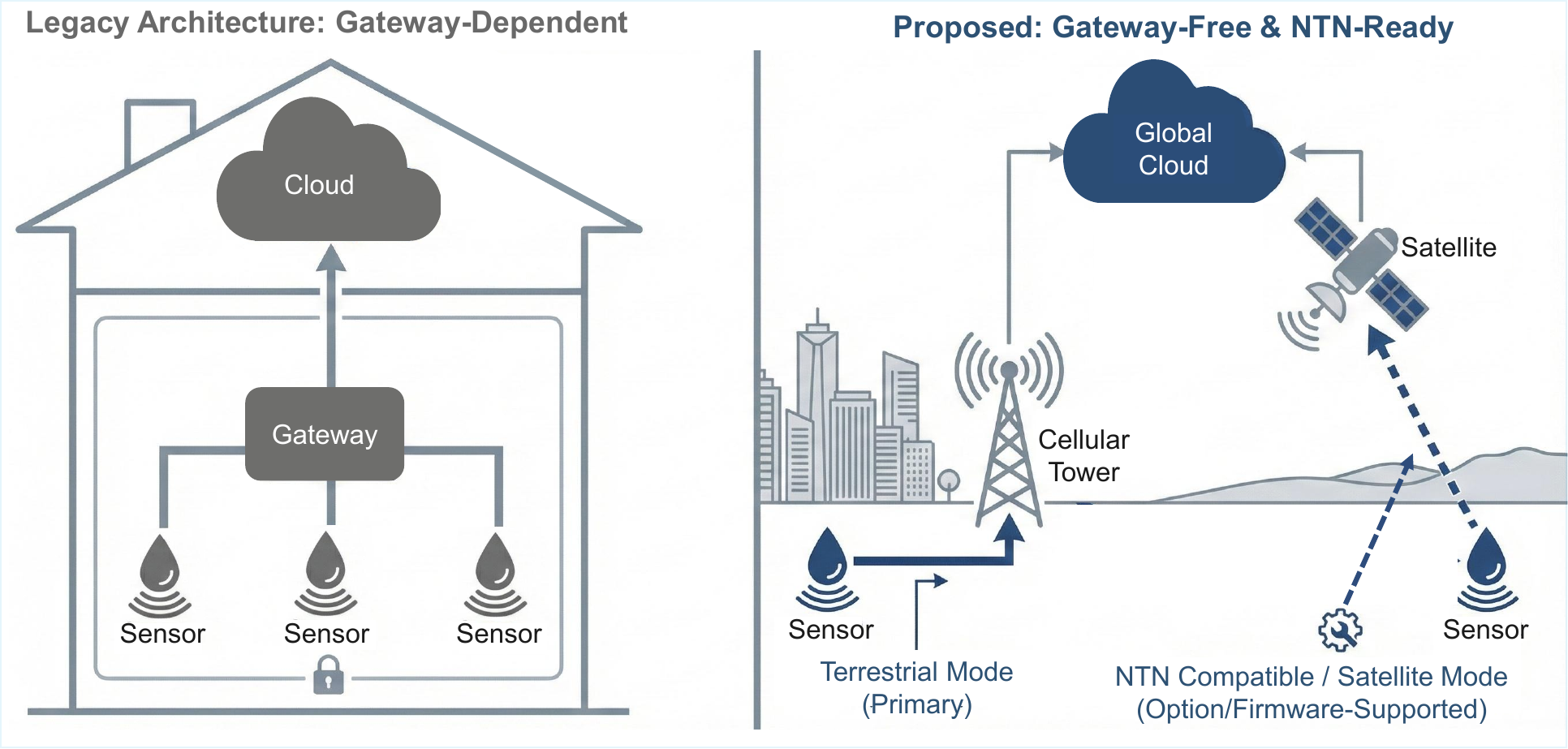} 
\caption{Evolution of leak detection architectures. The legacy approach (left) relies on local gateways and batteries, limiting scalability and increasing maintenance overhead. The proposed battery-free system (right) eliminates local infrastructure by connecting directly to terrestrial LTE-M towers, with firmware-supported compatibility for LEO satellites (NTN), enabling autonomous global monitoring.}
\label{fig:concept_diagram}
\end{figure*}

This paper addresses the limitations of existing battery-free leak-detection systems by introducing a fully battery- and gateway-free solution that leverages hydroelectric energy harvesting to power an LTE-M module for direct cloud communication, as shown in Fig.~\ref{fig:concept_diagram}. The proposed design uses a compartmentalized sensor architecture composed of materials that undergo electrochemical and electrophysical reactions upon water contact, allowing the leak itself to act as the energy source. Harvested energy is conditioned through a boost converter and stored in a supercapacitor, while a comparator-based control circuit ensures the LTE-M modem is enabled only when the voltage buffer exceeds a safe threshold. This prevents premature energy draw that could cause undervoltage lockout during network attachment or data transmission. The prototype integrates a commercial LTE-M platform (Thingy:91) for event-driven operation and can also be configured for non-terrestrial connectivity by constraining its operation to satellite-supported frequency bands. This design enables reliable communication in areas with limited terrestrial coverage and, to the best of our knowledge, represents the first demonstration of a fully battery-free system operating over LTE-M without a local gateway, paving the way for sustainable, maintenance-free IoT deployments at scale.

\noindent The main contributions of this work are as follows:
\begin{enumerate}
    \item Development of an event-powered hydroelectric (electrochemical) harvester integrated with a boost converter and supercapacitor buffer, enabling energy generation and storage directly from water exposure at the leak site.
    \item  A comparator-gated enable scheme that powers up the LTE-M modem only after the buffer exceeds a defined threshold, preventing brownouts during network attach and telemetry.
    \item Realization of a direct-to-cloud LTE-M communication link on a commercial module with event-driven firmware and MQTT-based telemetry.
    \item Demonstration of an optional non-terrestrial configuration that utilizes satellite-supported frequency bands for connectivity in remote or infrastructure-sparse environments.
    \item Comprehensive experimental validation of harvested energy, buffer behavior, attach performance, and message delivery success under event-triggered operation.
\end{enumerate}

 Collectively, these contributions advance the broader IoT field by demonstrating that energy harvested from the event itself can sustain cellular communication without batteries or gateways. This eliminates routine maintenance, simplifies deployment logistics, and establishes a pathway toward scalable, carrier-managed, and sustainable sensing networks across terrestrial and non-terrestrial domains.

The remainder of this paper is organized as follows. Section~\ref{sec:system_architecture} presents the overall system architecture and design rationale. Section~\ref{sec:communication_module} describes the LTE-M communication module, including hardware design, firmware workflow, power management circuitry, and non-terrestrial compatibility. Section~\ref{sec:experimental_results} outlines the experimental setup, measurement methods, results, and potential extensions to non-terrestrial operation. Finally, Section~\ref{sec:conclusion} concludes the paper with key insights and practical implications.

\section{System Architecture}
\label{sec:system_architecture}
\begin{figure*}[!t]
\centerline{\includegraphics[width=140mm]{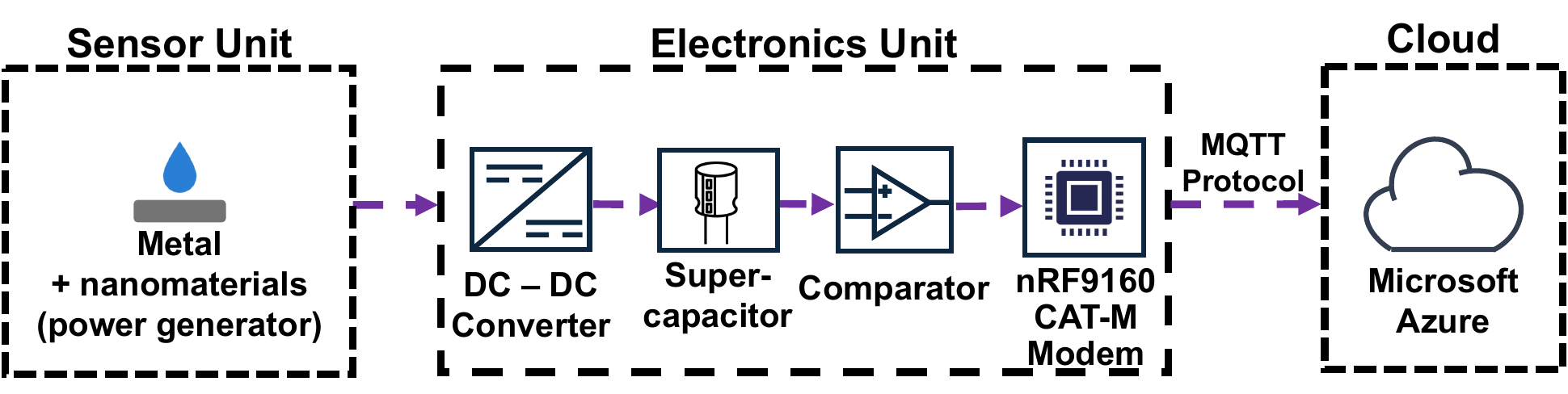}}
\caption{System architecture of the battery-free LTE-M system, where harvested energy powers the nRF9160 modem to transmit leak data to the cloud via MQTT.}
\label{fig:system_architecture}
\end{figure*}

Fig.~\ref{fig:system_architecture} presents an overview of the proposed battery-free IoT system. The design integrates energy harvesting, power management, and cellular communication into a compact sensing platform. The sensor autonomously generates power upon water exposure, which is regulated and buffered before enabling the LTE-M module for cloud transmission. The architecture ensures energy-aware operation without reliance on batteries or local gateways.

\subsection{Sensor Unit}

The sensor unit is designed to harvest energy from water-triggered electrochemical reactions using a nanomaterial-based structure. This mechanism forms the energy backbone of the system, enabling completely battery-free operation. The sensor consists of a layered material stack comprising carbon nanofibers (CNFs) mixed with metallic chloride, sandwiched between reactive metal electrodes, typically aluminum (Al) and magnesium (Mg)~\cite{feng2020high}. Unlike conventional lithium-based power sources that pose disposal hazards, these abundant, non-toxic materials align with the sustainability goals of large-scale environmental monitoring. Upon contact with water, the system initiates electrochemical and electrophysical reactions, generating electrical energy without any external power source.

In prior work involving BLE and LoRa-based systems, this sensor architecture demonstrated the feasibility of harvesting sufficient energy for low-power communication~\cite{Rouhi2024, nepal_Lora}. Specifically, in the LoRa implementation, a single 55~mm diameter sensor produced a peak Open Circuit Voltage (OCV) of approximately \SI{1.65}{\volt}, a Short Circuit Current (SCC) of \SI{200}{\milli\ampere} to \SI{250}{\milli\ampere}, peaking around \SI{500}{\milli\ampere}. This output was adequate for powering a LoRa module using a DC-DC boost converter and a \SI{100}{\milli\farad} supercapacitor. The power demands of LoRa are moderate, so the current drop caused by stepping up \SI{1.65}{\volt} to \SI{5}{\volt} did not significantly hinder performance.

However, LTE-M communication requires significantly more power, both in terms of higher minimum operating voltage (\( \geq \SI{3.2}{\volt} \)) and initial startup currents exceeding \SI{250}{\milli\ampere}~\cite{nordic2022thingy91,Lauridsen2018}. In such high-demand scenarios, directly stepping up from \SI{1.65}{\volt} results in substantial current losses, leading to extended capacitor charging times and unreliable module activation. To address this, we introduce a system architecture optimization via a series-compartment sensor configuration.

\begin{figure}[!t]
\centerline{\includegraphics[width=80mm]{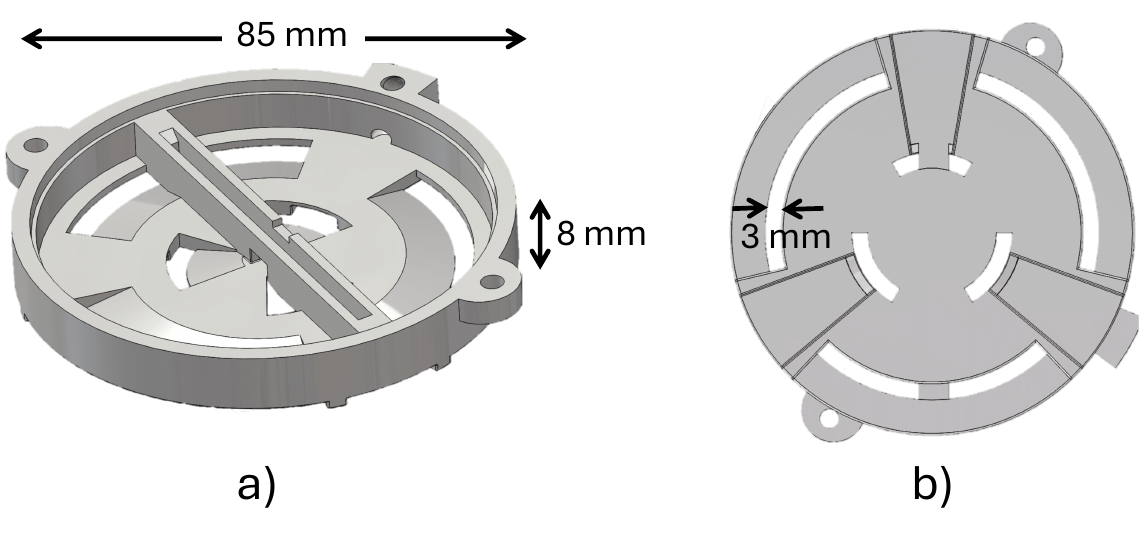}}
\caption{CAD model of the sensor enclosure: (a) Isometric view showing the internal compartments designed to hold electrode and electrolyte materials; (b) Bottom view illustrating inlet channels that allow water to enter and trigger the electrochemical reaction for energy generation.}
\label{fig:sensor_unit}
\end{figure}

In this new sensor configuration, two electrochemical compartments, each with the same material composition, are connected in an internal series arrangement, as shown in Fig.~\ref{fig:sensor_unit}. This structural modification increases the OCV without compromising the current output. By generating a higher voltage natively, this series-compartment design reduces the voltage step-up ratio required by the DC-DC converter, thereby minimizing conversion losses and enhancing the efficiency of energy transfer to the supercapacitor. Despite the added structural complexity, the voltage gain significantly improves overall power management and supports higher-power communication protocols such as LTE-M. 

Additionally, the electrode geometry incorporates a 3~mm inlet channel width (as shown in Fig.~\ref{fig:sensor_unit}b) to prevent false triggers caused by ambient humidity or condensation. This ensures that the electrochemical reaction is initiated only by the accumulation of liquid water sufficient to bridge the contacts, thereby enhancing detection reliability in damp environments.

\subsection{Power Management Unit}

To bridge the gap between the low-voltage, variable output of the sensor and the high, stable power requirements of the LTE-M module, a dedicated power management unit is integrated into the system. This unit is responsible for conditioning, storing, and regulating the harvested energy before it is delivered to the communication hardware. It consists of three main components: a boost converter to step-up the sensor output to a usable voltage level, a supercapacitor to store the accumulated energy, and a comparator-based circuit to control the timing of power delivery to the LTE-M board. Together, these elements enable intermittent energy harvesting to be converted into short bursts of high-power operation, ensuring reliable and autonomous system functionality.

\subsubsection{Boost Converter}
\label{subsec:boost_converter}

The energy harvested from the sensor produces a peak voltage in the range of approximately 2.7~V with average of 1.6~V, which is insufficient to directly power LTE-M modules that require a minimum of 3.2~V~\cite{nordic2022thingy91}. To address this, a step-up DC-DC boost converter is employed to elevate the voltage to a regulated 5~V, enabling efficient energy storage and downstream operation.

The system utilizes the ME2108A50, a compact step-up converter that employs a Pulse Frequency Modulation (PFM) topology~\cite{ME2108_datasheet}. Unlike micro-watt harvesters (e.g., RF or piezoelectric) that rely on Maximum Power Point Tracking (MPPT) to extract energy from high-impedance sources, the electrochemical sensor generates sufficient short-circuit current to drive a standard boost converter directly. As a result, the PFM topology was selected to maintain regulation even when the sensor voltage sags under heavy load. This choice also avoids the quiescent current overhead and tracking latency associated with complex MPPT circuits.

\begin{figure}[!t]
\centerline{\includegraphics[width=80mm]{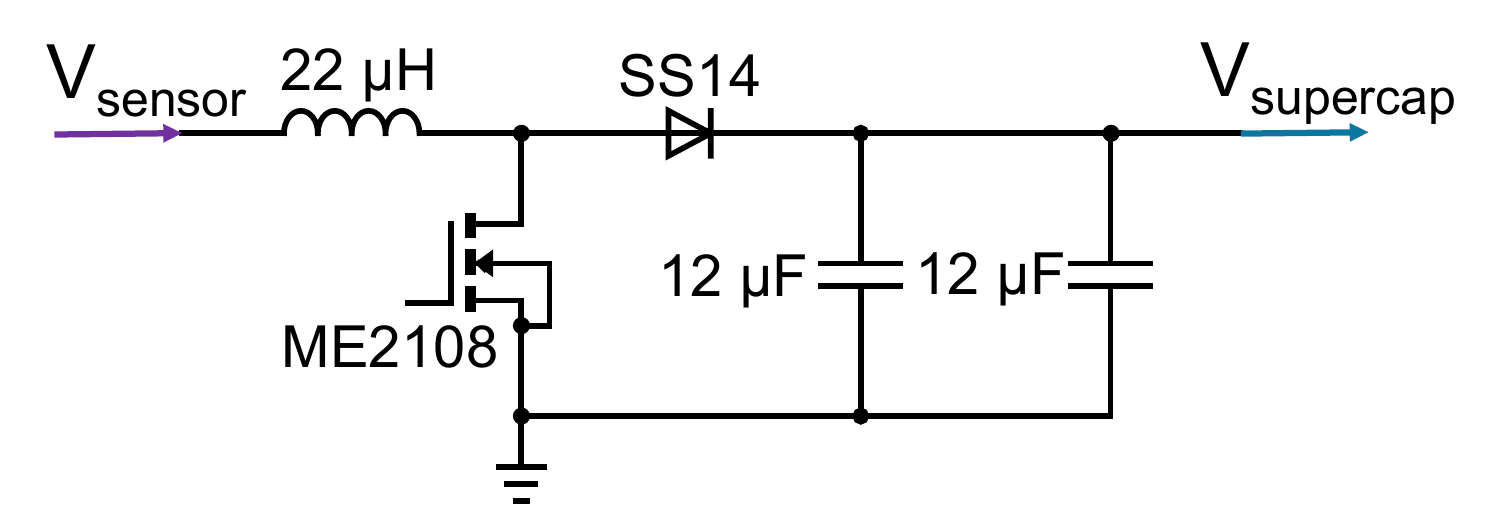}}
\caption{Schematic of the DC-DC boost converter used to step up the sensor output voltage to 5~V, which gets stored in a supercapacitor for powering the LTE-M module.}
\label{fig:boost_converter}
\end{figure}

The converter operates with an internal oscillator frequency of 180~kHz, balancing conversion efficiency with component size~\cite{ME2108_datasheet}. To ensure stable operation, the external circuit includes a \SI{22}{\micro\henry} inductor selected to minimize ripple current and avoid saturation during peak loads, alongside a low forward-voltage Schottky diode (SS14) to reduce freewheeling losses. Two \SI{12}{\micro\farad} capacitors are placed at the output to smooth voltage fluctuations during switching, as shown in Fig.~\ref{fig:boost_converter}.

The ME2108A50 supports start-up voltages as low as \SI{0.9}{\volt}~\cite{ME2108_datasheet}. This low-voltage start-up capability is essential for this intermittently powered system, allowing the converter to begin harvesting energy in the earliest stages of a leak event. The regulated output is directed to a supercapacitor, which accumulates energy until it reaches the threshold required to trigger LTE-M transmission.

\subsubsection{Load Isolation via Comparator}

In energy-harvesting systems with limited input power, premature connection of the load can lead to undervoltage operation and system instability~\cite{Sudevalayam2011}. This is especially critical for LTE-M communication modules, which require a high startup current. If the load connects prematurely while the supercapacitor is charging, the discharge rate often exceeds the recharge rate. This creates a net energy loss, preventing the system from ever reaching the voltage required for reliable operation~\cite{Dar2024,Ban2013}.

\begin{figure}[!t]
\centerline{\includegraphics[width=80mm]{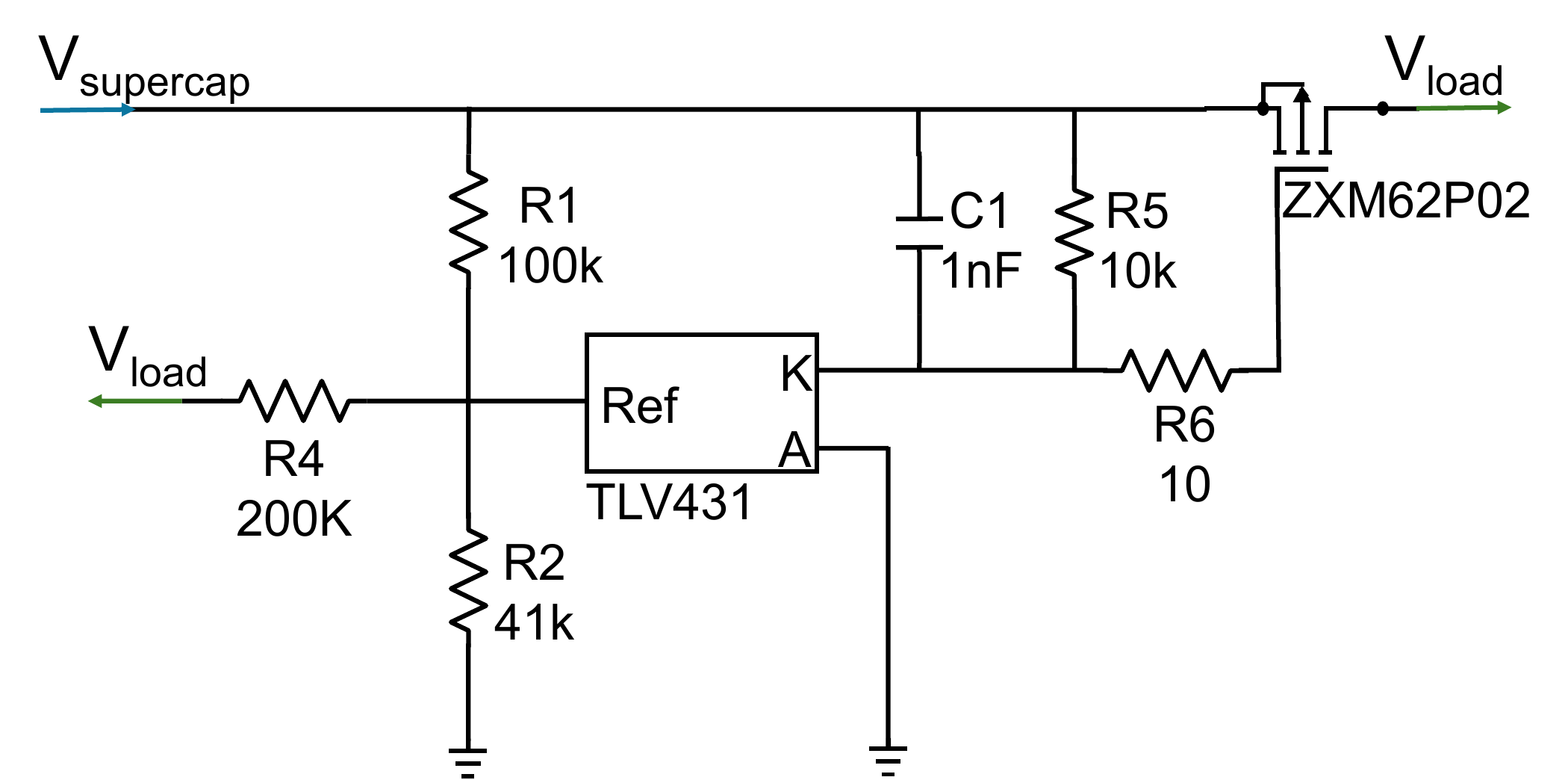}}
\caption{ Load isolation circuit using a TLV431-based comparator and P-channel MOSFET. The resistor network ($R_1$, $R_2$, $R_4$) sets the hysteretic voltage thresholds, enabling the load when the supercapacitor voltage exceeds $V_{\text{on}}$ and disconnecting it below $V_{\text{off}}$.}
\label{fig:comparator_circuit}
\end{figure}

To address this, the proposed architecture includes a comparator-based load isolation circuit that ensures the LTE-M module remains disconnected until sufficient energy has been accumulated. Fig.~\ref{fig:comparator_circuit} shows the schematics of the load isolation circuit. It uses a TLV431 precision shunt regulator configured as a comparator, along with a ZXM62P02 P-channel MOSFET that acts as a high-side switch~\cite{tlv431_datasheet, zxm62p02e6_datasheet}. A resistive network consisting of $R_1$, $R_2$, and $R_4$ monitors the supercapacitor voltage $V_{\text{cap}}$ and applies a scaled version to the TLV431 reference pin.

When $V_{\text{cap}}$ exceeds a predefined threshold, the reference pin crosses the TLV431's internal reference voltage ($V_{\text{ref}} = \SI{1.24}{\volt}$), causing it to sink current and turn on the MOSFET. This connects the LTE-M module to the supply, enabling data transmission. The load is only enabled when the capacitor has charged to a minimum of \SI{4.87}{\volt}, ensuring sufficient energy for a complete transmission cycle.

To prevent oscillations and provide controlled disconnection, the circuit includes hysteresis. Once enabled, the comparator maintains the MOSFET in the on state until the capacitor discharges down to a lower threshold set at \SI{3.2}{\volt}. Below this voltage, the TLV431 disables the MOSFET, isolating the LTE-M module and allowing the capacitor to recharge without being drained by the load. This controlled two-threshold behavior guarantees efficient cycling between energy accumulation and load activation. Furthermore, the resistive network values ($R_1, R_2, R_4$) were specifically selected in the $100~k\Omega$ range to minimize the static quiescent current drawn by the comparator circuit. This ensures that the control logic consumes negligible power ($< 40~\mu A$) relative to the sensor's generation rate, preventing the isolation circuit itself from becoming a significant parasitic load during the charging phase.

The turn-on and turn-off voltages are defined by the resistor network and can be expressed as:
\begin{align}
V_{\text{on}} &= V_{\text{ref}} \left( 1 + \frac{R_1}{R_2 \parallel R_4} \right), \label{eq:von} \\
V_{\text{off}} &= V_{\text{ref}} \left( 1 + \frac{R_1 \parallel R_4}{R_2} \right), \label{eq:voff}
\end{align}
The separation between $V_{\text{on}}$ and $V_{\text{off}}$ introduces hysteresis, which prevents chattering around the threshold and provides a clear margin between activation and disconnection.

The operation can be divided into three distinct phases:
\begin{itemize}
    \item \textit{Charging Phase:} When $V_{\text{cap}} < V_{\text{on}}$, the MOSFET remains off, isolating the LTE-M module and allowing uninterrupted capacitor charging.
    \item \textit{Activation Phase:} Once $V_{\text{cap}} \geq V_{\text{on}}$, the comparator enables the MOSFET, powering the LTE-M module for data transmission.
    \item \textit{Disconnection Phase:} As the module draws current and the capacitor discharges, the comparator turns off the MOSFET when $V_{\text{cap}}$ drops below $V_{\text{off}}$, re-isolating the load and restarting the charging cycle.
\end{itemize}

\begin{figure}[!t]
\centerline{\includegraphics[width=80mm]{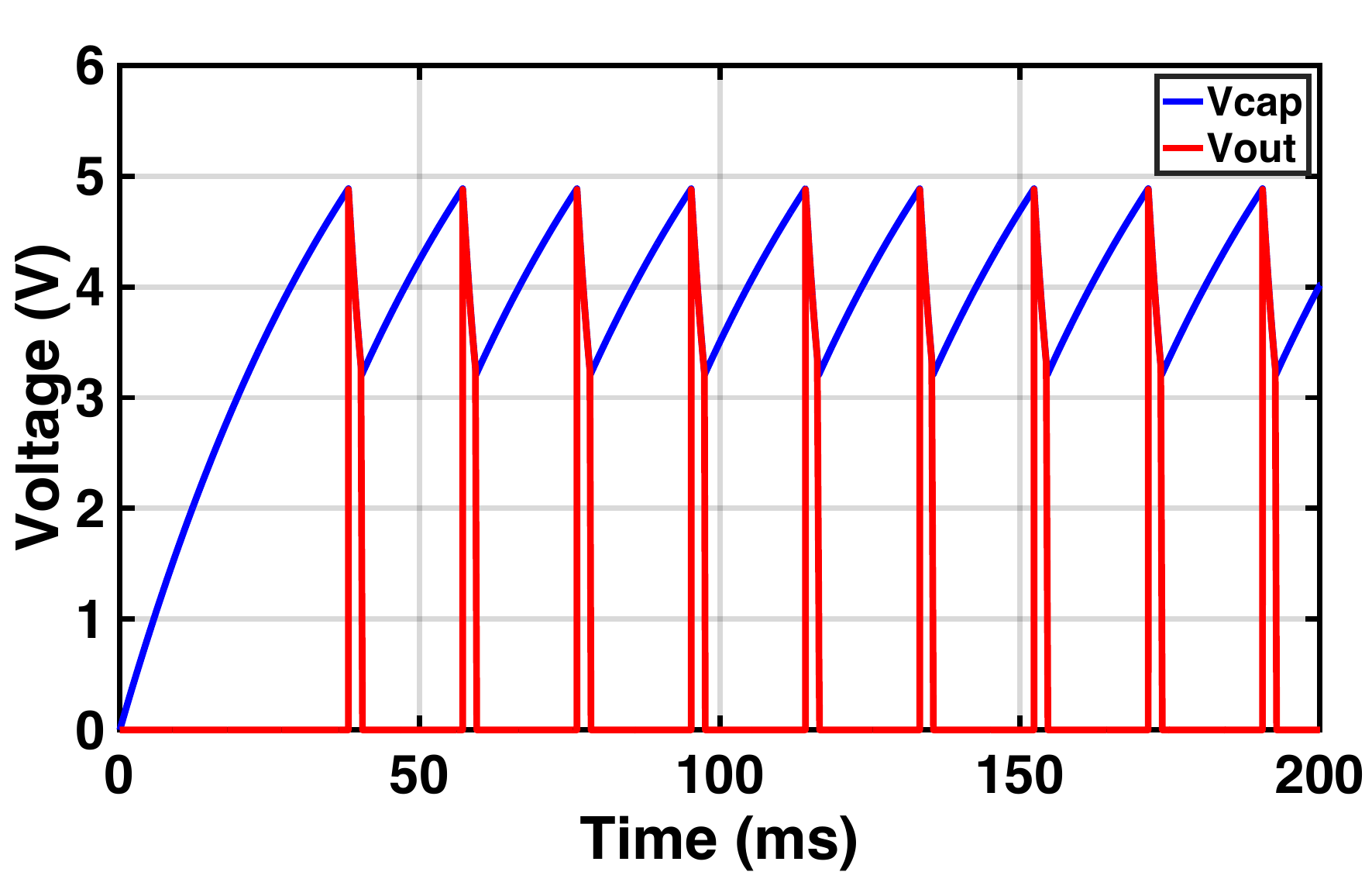}}
\caption{LTspice simulation showing $V_{\text{cap}}$ rising with energy harvesting and $V_{\text{out}}$ toggling as the load is connected and disconnected based on comparator thresholds.}
\label{fig:comparator_simulated}
\end{figure}

To verify the switching functionality of the circuit, Fig.~\ref{fig:comparator_simulated} shows the simulated behavior using LTspice, where a simplified model of the energy harvesting circuit was created using diode rectifiers and a \SI{47}{\micro\farad} capacitor to reduce simulation time. Since the goal is to validate the turn-on and turn-off thresholds of the comparator circuit, the reduced capacitor size has no impact on qualitative behavior. The graph shows $V_{\text{cap}}$ and $V_{\text{out}}$ over time. Initially, the capacitor charges with the load isolated. Once the voltage reaches $V_{\text{on}}$, the comparator activates the MOSFET, enabling the load and causing a drop in $V_{\text{cap}}$. When the voltage falls below $V_{\text{off}}$, the load is disconnected and the capacitor resumes charging—repeating the cycle predictably.

\subsubsection{Supercapacitor Storage}

The supercapacitor serves as the primary energy buffer in the system, bridging the mismatch between the low, intermittent output of the energy-harvesting sensor and the high instantaneous power demand of the LTE-M communication module~\cite{Subasinghage2024,Zhong2017}. Its role is to accumulate energy over time and release it in bursts during energy-intensive events such as network registration and data transmission.

To determine the required capacitance, we first consider the energy needed to support a complete communication cycle: initial network search, idle standby, and one subsequent transmission. Based on the power consumption profile characterized in Section~\ref {subsec:power consumption analysis} (Table~\ref{tab:ltem_energy}), the total measured energy consumption across these phases is:

\[
E_{\text{meas}} = 2.15~\text{J} + 0.984~\text{J} + 0.3523~\text{J} = 3.49~\text{J}
\]

Given that the system uses a boost converter with approximately 75\% efficiency (Section \ref{subsec:boost_converter_efficiency}, the actual energy that must be supplied by the capacitor is:

\begin{equation}
E_{\text{cap}} = \frac{E_{\text{meas}}}{0.75} = \frac{3.49}{0.75} \approx 4.65~\text{J}
\end{equation}

Next, we compute the required capacitance based on the usable energy between the comparator-controlled voltage thresholds: $V_{\text{on}} = \SI{4.87}{\volt}$ and $V_{\text{off}} = \SI{3.25}{\volt}$. The energy available from a capacitor discharging over this range is:

\begin{equation}
\Delta E = \frac{1}{2} C \left( V_{\text{on}}^2 - V_{\text{off}}^2 \right) = \frac{1}{2} C (23.72 - 10.56) = 6.58\,C
\end{equation}

Solving for the required capacitance:

\begin{equation}
C = \frac{E_{\text{cap}}}{6.58} = \frac{4.65}{6.58} \approx \SI{0.71}{\farad}
\label{eq:cap_calc}
\end{equation}

While this analysis shows that a 0.71~F capacitor is theoretically sufficient, practical considerations suggest using a higher value. In reality, the energy-harvesting sensor continues to deliver power during capacitor discharge, particularly during the idle phase. This background charging allows the capacitor to partially recover between transmission cycles, improving its ability to sustain multiple beacon events (Section~\ref{sec:activation_and_communication}).

\begin{figure}[!t]
\centerline{\includegraphics[width=80mm]{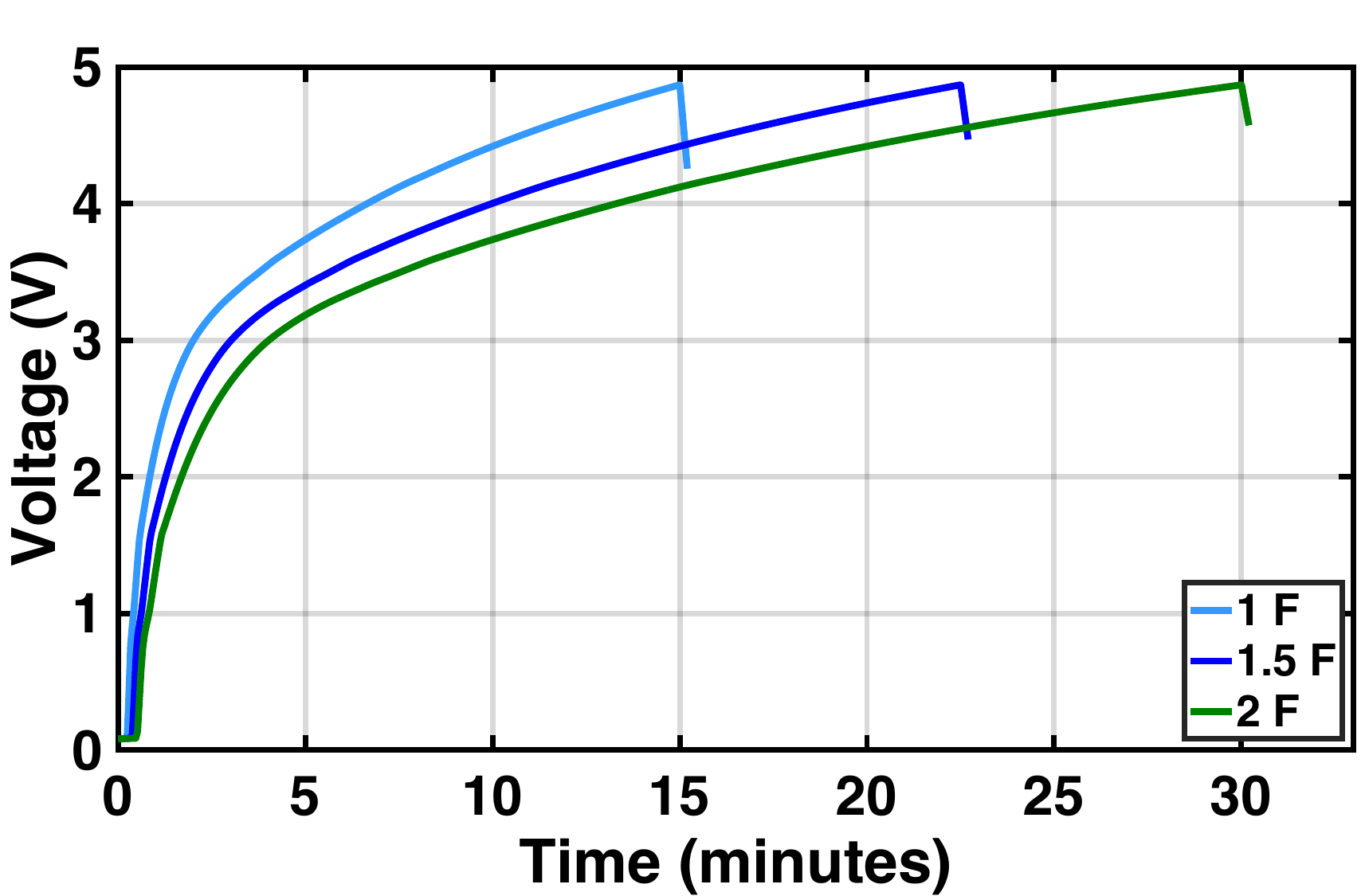}}
\caption{Supercapacitor voltage profiles for 1~F, 1.5~F, and 2~F capacitors during charging and discharging, illustrating the impact of capacitance on charge time and voltage drop behavior during LTE-M load activation.}
\label{fig:cap_charging_discharging}
\end{figure}

To ensure reliable operation across a range of leak scenarios, a 1.5~F supercapacitor was selected for all subsequent experiments. Fig.~\ref{fig:cap_charging_discharging} compares the voltage profiles of 1~F, 1.5~F, and 2~F capacitors during charging and discharging. The 1~F capacitor shows a shorter charge time but experiences a comparatively higher voltage drop during the LTE-M startup phase, which can jeopardize communication reliability. Conversely, a 2~F capacitor offers better voltage stability but increases charging time. The 1.5~F configuration provides a practical middle ground, achieving reliable activation while maintaining a reasonable charging time.

\section{LTE-M Communication Module}
\label{sec:communication_module}

\subsection{Hardware}

\begin{figure}[!t]
\centerline{\includegraphics[width=50mm]{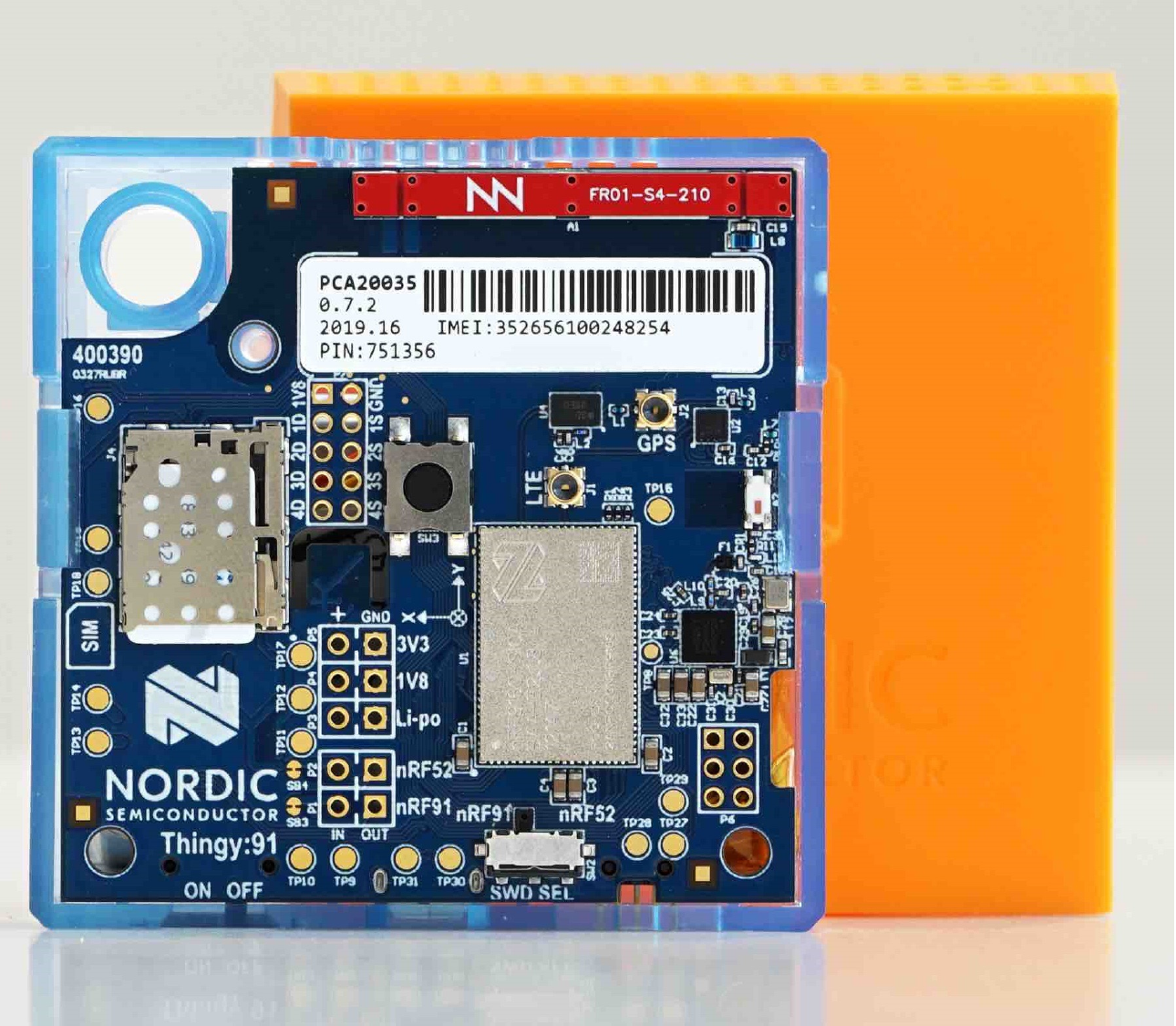}}
\caption{Nordic Thingy:91 module used for direct-to-cloud LTE-M communication in the battery-free leak detection system.}
\label{fig:thingy91}
\end{figure}

The LTE-M communication system is built around the Nordic Thingy:91 evaluation board, which integrates the nRF9160 SiP~\cite{nordic2022thingy91, nordic_nrf9160}, as shown in Fig~\ref{fig:thingy91}. This integrated module combines a cellular modem with an Arm Cortex-M33 processor, offering native support for both LTE Cat-M1 (LTE-M) and NB-IoT. For this work, only LTE-M is utilized, due to its wider network availability in North America.

The nRF9160 supports LTE-M operation across a broad range of E-UTRA frequency bands, including Bands 1, 2, 3, 4, 5, 8, 12–14, 17–20, 25, 26, 28, and 66~\cite{nordic2022thingy91, nordic_nrf9160}. This wide band compatibility enhances its suitability for global deployment and allows integration with a variety of cellular carriers. The modem operates in Type B half-duplex FDD mode and is compliant with 3GPP LTE Release 13, ensuring compatibility with modern cellular IoT infrastructures.

To align with the low-power, battery-free operation of the system, several modifications are made to the Thingy:91 board. The onboard GNSS, integrated environmental sensors, and LEDs are disabled to minimize peripheral power draw. Furthermore, the Li-Po battery is removed entirely, and all power is delivered through the energy-harvesting unit via the supercapacitor buffer. 

\subsection{Firmware}

The firmware for the LTE-M communication module is developed using the nRF Connect SDK, which builds upon the Zephyr Real-Time Operating System (RTOS). It is compiled with built-in support for secure boot and image verification, and flashed to the Thingy:91 development board using the nRF Connect for Desktop utility. The resulting image is signed and encrypted to enable secure boot with image authentication, ensuring firmware integrity throughout the device lifecycle.

The system follows an event-driven architecture centered around energy availability. Upon activation, triggered externally when the supercapacitor voltage exceeds \SI{4.87}{\volt} as determined by the comparator, the firmware boots the application processor, initializes the LTE-M modem, and begins the network connection procedure. LTE Cat-M1 functionality is enabled, while NB-IoT and GNSS are disabled to reduce radio scan time and focus power usage solely on the required uplink task.

After successful network registration, the firmware sends a telemetry beacon to a cloud endpoint using the MQTT protocol (QoS 0) via the Azure IoT Hub SDK. The message includes event status and timestamp. To ensure data integrity and confidentiality suitable for critical infrastructure deployment, the communication link is secured using Transport Layer Security (TLS 1.2). The modem handles the encryption offload, establishing a secure tunnel to the cloud before the MQTT payload is transmitted, thereby protecting the device from spoofing or man-in-the-middle attacks. Communication with the modem is conducted via a socket-based AT command interface.

Following transmission, the firmware enforces a fixed idle mode of two minutes, during which the processor remains inactive. This delay allows the supercapacitor to partially recharge from the energy-harvesting sensor, while the modem remains idle. In the current implementation, this sleep routine is handled entirely by firmware and occurs independently of comparator-based voltage thresholds. The complete firmware logic is illustrated in Fig.~\ref{fig:firmware_flowchart}.

% TikZ styles
\tikzstyle{startstop} = [rectangle, rounded corners, minimum width=2cm, minimum height=0.8cm, text centered, draw=red, fill=red!30, font=\fontsize{8}{12}\selectfont]
\tikzstyle{process} = [rectangle, minimum width=2cm, minimum height=0.8cm, text centered, draw=orange, fill=orange!30, font=\fontsize{8}{12}\selectfont]
\tikzstyle{arrow} = [thick,->,>=stealth]

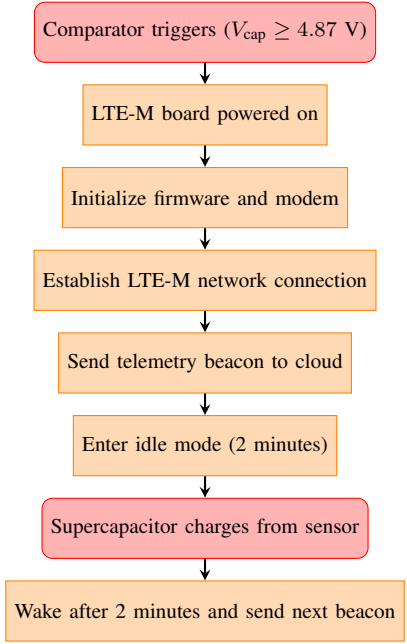
\begin{figure}[!t]
\begin{center}
\begin{tikzpicture}[node distance=1.1cm]

% Nodes
\node (start) [startstop] {Comparator triggers ($V_\text{cap} \geq 4.87$ V)};
\node (powerup) [process, below of=start] {LTE-M board powered on};
\node (init) [process, below of=powerup] {Initialize firmware and modem};
\node (connect) [process, below of=init] {Establish LTE-M network connection};
\node (send) [process, below of=connect] {Send telemetry beacon to cloud};
\node (sleep) [process, below of=send] {Enter idle mode (2 minutes)};
\node (charge) [startstop, below of=sleep] {Supercapacitor charges from sensor};
\node (wake) [process, below of=charge] {Wake after 2 minutes and send next beacon};

% Arrows
\draw [arrow] (start) -- (powerup);
\draw [arrow] (powerup) -- (init);
\draw [arrow] (init) -- (connect);
\draw [arrow] (connect) -- (send);
\draw [arrow] (send) -- (sleep);
\draw [arrow] (sleep) -- (charge);
\draw [arrow] (charge) -- (wake);

\end{tikzpicture}
\end{center}
\caption{Firmware flowchart of the LTE-M sensor node. The system activates once the supercapacitor voltage exceeds the comparator threshold, performs a single telemetry transmission, enters deep sleep while the capacitor continues charging, and wakes after two minutes to repeat the process.}
\label{fig:firmware_flowchart}
\end{figure}

\subsection{Power Management}

The nRF9160 SiP supports multiple power-saving features, including Power Saving Mode (PSM) and extended Discontinuous Reception (eDRX), which are designed to reduce average current consumption in cellular IoT applications. While these features are supported by the hardware and available in the development environment, they are intentionally left disabled in this work to evaluate system performance under worst-case power conditions.

This means the LTE-M modem remains in connected idle mode following each transmission, where it continues to draw a relatively high current due to periodic network paging and lack of deep sleep. This scenario is used to validate the strength and robustness of the energy-harvesting system, demonstrating that even under unfavorable conditions with no modem-level power optimization, the system is capable of completing repeated charge-transmit cycles without external power or batteries.

Although enabling PSM would reduce standby current to the microampere range and significantly extend the number of transmission cycles per charge, it is omitted here to stress-test the hardware design. Similarly, eDRX is not used, as the application is uplink-only and does not require periodic downlink communication. Disabling both features ensures that the device operates continuously in a high-power idle state, representing the most demanding operational profile.

These choices emphasize the energy-harvesting system's ability to support the LTE-M module without auxiliary power management enhancements. Since the system is able to operate reliably under worst-case power conditions, enabling power-saving modes would significantly enhance performance. If PSM were enabled, reducing the floor current to $\sim\SI{3}{\micro\ampere}$ compared to the milliampere-range idle current observed here, the energy budget required for the idle phase would drop by orders of magnitude. This would theoretically eliminate parasitic discharge from the supercapacitor between transmission events, accelerating voltage recovery and maximizing the total number of beacon transmissions achievable in a single charge cycle.

\subsection{Non-terrestrial Compatibility}

While the primary implementation of the system targets terrestrial LTE-M networks, certain deployment scenarios, such as rural, offshore, or disaster-affected regions, may lack reliable cellular coverage. To extend the applicability of the proposed battery-free leak detection node to these environments, the system was configured for non-terrestrial operation through a modification at the modem level. This configuration does not require any new hardware or alterations to the power management structure. Instead, it constrains the LTE-M modem to operate within satellite-supported frequency bands, allowing the existing architecture to maintain functionality beyond terrestrial network limits.

In the Thingy:91 module, the LTE-M modem is by default programmed to support a wide range of frequency bands, covering standard terrestrial cellular networks. However, emerging D2D satellite services, such as SpaceX's Starlink and Amazon's Project Kuiper, leverage specific existing cellular frequencies (e.g., the uplink of Band~2, 1850–1910~MHz) to serve unmodified UE~\cite{Vaezi2022,Moges2023}. To maintain a reliable link with moving LEO nodes in these bands, the system architecture supports the 3GPP Release 17 NTN protocol stack, which abstracts physical link challenges by utilizing GNSS positioning data to actively apply frequency pre-compensation and timing advance adjustments directly within the firmware.

To evaluate the system's ability to operate within this specific satellite-compatible spectrum, the modem configuration was modified using the \texttt{AT\%XCBAND} command to explicitly restrict operation to Band~2. Under this frequency-constrained configuration, the node successfully initialized and sustained operation, demonstrating the hardware and energy feasibility of satellite-supported LTE-M operation within the same energy-managed architecture.

The proposed configuration enables the node to operate under both terrestrial and non-terrestrial network conditions. In standard environments, the device connects to conventional LTE-M base stations, while in regions lacking terrestrial coverage, it can maintain connectivity through LEO satellites. The firmware can be configured to automatically switch between terrestrial and satellite-supported modes based on network availability, ensuring continuous connectivity and enhancing overall system resilience.

All system parameters, including comparator-based power gating and supercapacitor energy buffering, remain unchanged. Peripheral components such as GNSS, onboard sensors, and indicator LEDs remain disabled to minimize current draw and preserve harvested energy. The firmware-level configurability extends the node’s operational domain from connected environments to a ubiquitous sensing platform, enabling autonomous, event-driven monitoring in diverse and remote locations globally while preserving its core battery-free energy harvesting model.

\section{Experimental Results}
\label{sec:experimental_results}

\subsection{Network Context and Measurement Setup}
All experiments were performed using the LTE-M interface of the nRF9160 System-in-Package on the Nordic Thingy:91 development platform. The modem was allowed to automatically select the serving cell and operating band based on network availability; no manual band locking was applied. Serving-cell metrics were obtained before and after each trial through the AT command interface (\texttt{AT\%XMONITOR} for RSRP, RSRQ, and SINR; \texttt{AT+COPS?} for PLMN identification). The measured parameters and environmental conditions are summarized in Table~\ref{tab:radio_conditions}. These values represent a typical indoor LTE-M connection under moderate signal quality and were used to interpret communication timing and energy results.

\begin{table*}[t]
\centering
\begin{threeparttable} % Wraps the tabular
    \caption{Radio conditions and network context during experiments}
    \label{tab:radio_conditions}
    \begin{tabular}{lcc}
    \toprule
    Metric & Value & Notes \\
    \midrule
    Operator (PLMN) & Rogers Wireless (302–720) & APN: \texttt{ltemobile.apn} \\
    Serving EARFCN / PCI & Auto-selected & Not explicitly logged (auto-band selection) \\
    RSRP (dBm) & $-97.8 \pm 3.9$ & Range: $[-104, -92]$, $N{=}10$ \\
    RSRQ (dB) & $-9.3 \pm 1.3$ & Range: $[-12.0, -7.5]$ \\
    SINR (dB) & $11.5 \pm 2.8$ & Range: $[6.5, 16.0]$ \\
    Environment & Indoor lab & $\sim$300~m from base station \\
    Temperature & 22~\textdegree C & Controlled room conditions \\
    \bottomrule
    \end{tabular}
    \begin{tablenotes}
        \footnotesize
        \item \textit{Definitions:} PLMN: Public Land Mobile Network; APN: Access Point Name; EARFCN: E-UTRA Absolute Radio Frequency Channel Number; PCI: Physical Cell Identity; RSRP: Reference Signal Received Power; RSRQ: Reference Signal Received Quality; SINR: Signal-to-Interference-plus-Noise Ratio.
    \end{tablenotes}
\end{threeparttable}
\end{table*}

\subsection{Energy Generation and Storage}

To characterize the performance of the dual-compartment sensor, a series of experiments were conducted to measure its electrical output under controlled conditions. The sensor was activated by placing it in a Petri dish with a shallow water depth of 1~mm—chosen deliberately to replicate realistic leak conditions, where surface water spreads laterally rather than pooling in volume. This approach better captures the sensor’s sensitivity to early-stage leaks, as water depth is a more practical metric than volume in real-world applications.

Separate measurements were taken for OCV and SCC using an Agilent 34411A digital multimeter (DMM)~\cite{keysight34410A34411A}. The OCV measurement provides insight into the maximum voltage the sensor can generate in the absence of a load, while the SCC measurement provides the peak current the sensor can deliver when shorted~\cite{Zhang2018}. These two metrics together are important for evaluating energy harvesting capability and compatibility with wireless protocols such as LTE-M.

\begin{figure}[!t]
\centerline{\includegraphics[width=90mm]{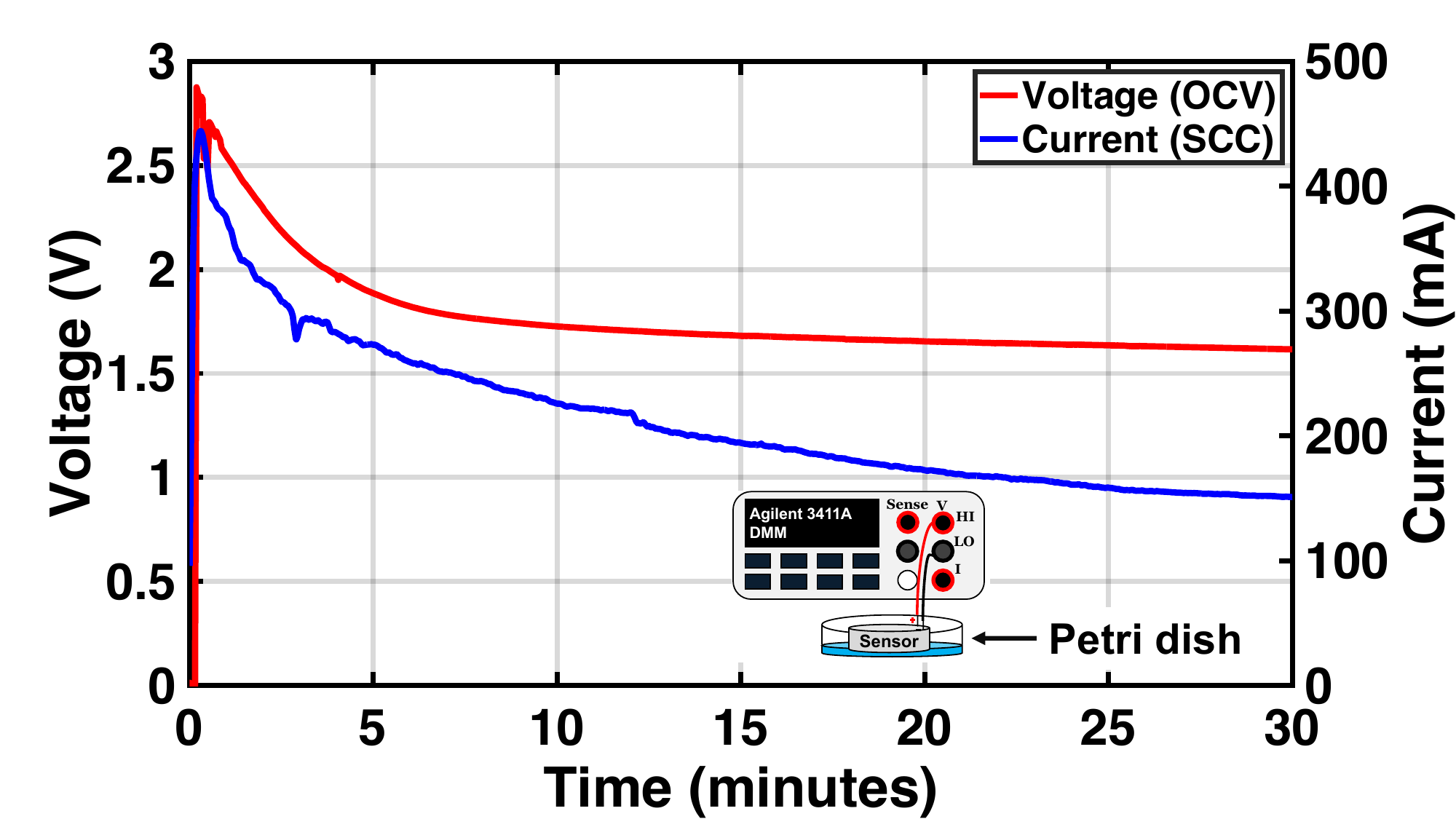}}
\caption{OCV (left y-axis) and SCC (right y-axis) output of the sensor over time after water exposure. Inset: experimental setup using an Agilent 34411A DMM and a Petri dish for activation.}
\label{fig:OCV_SCC}
\end{figure}

As illustrated in Fig.~\ref{fig:OCV_SCC}, the sensor produces a peak OCV of approximately 2.7~V shortly after water exposure, which gradually stabilizes to around 1.6~V over a 30-minute period. Meanwhile, the SCC exceeds 450~mA initially, stabilizing at approximately 150~mA. These results demonstrate significant energy harvesting potential, especially in comparison to earlier single-compartment versions of the sensor~\cite{Rouhi2024}.

The improvement in voltage is attributed to the series configuration of the two electrochemical compartments, which raises the overall potential difference across the terminals. However, it is important to note that the increase is not perfectly linear; the total energy output is limited by the amount of water available, and since the two compartments draw from the same source, their electrochemical reactions are not entirely independent. Nonetheless, the architectural enhancement provides a practical voltage gain without compromising current levels, thereby making the sensor more suitable for powering energy-intensive LTE-M communication modules.

\subsection{Activation and Communication}
\label{sec:activation_and_communication}

To evaluate the complete system behavior, the energy harvesting sensor, power management circuitry, supercapacitor, and LTE-M module were fully integrated and tested in a closed-loop configuration. The voltage across the 1.5~F supercapacitor, $V_{\text{cap}}$, was continuously monitored using a DMM to analyze system dynamics during autonomous communication cycles.

\begin{figure}[!t]
\centerline{\includegraphics[width=80mm]{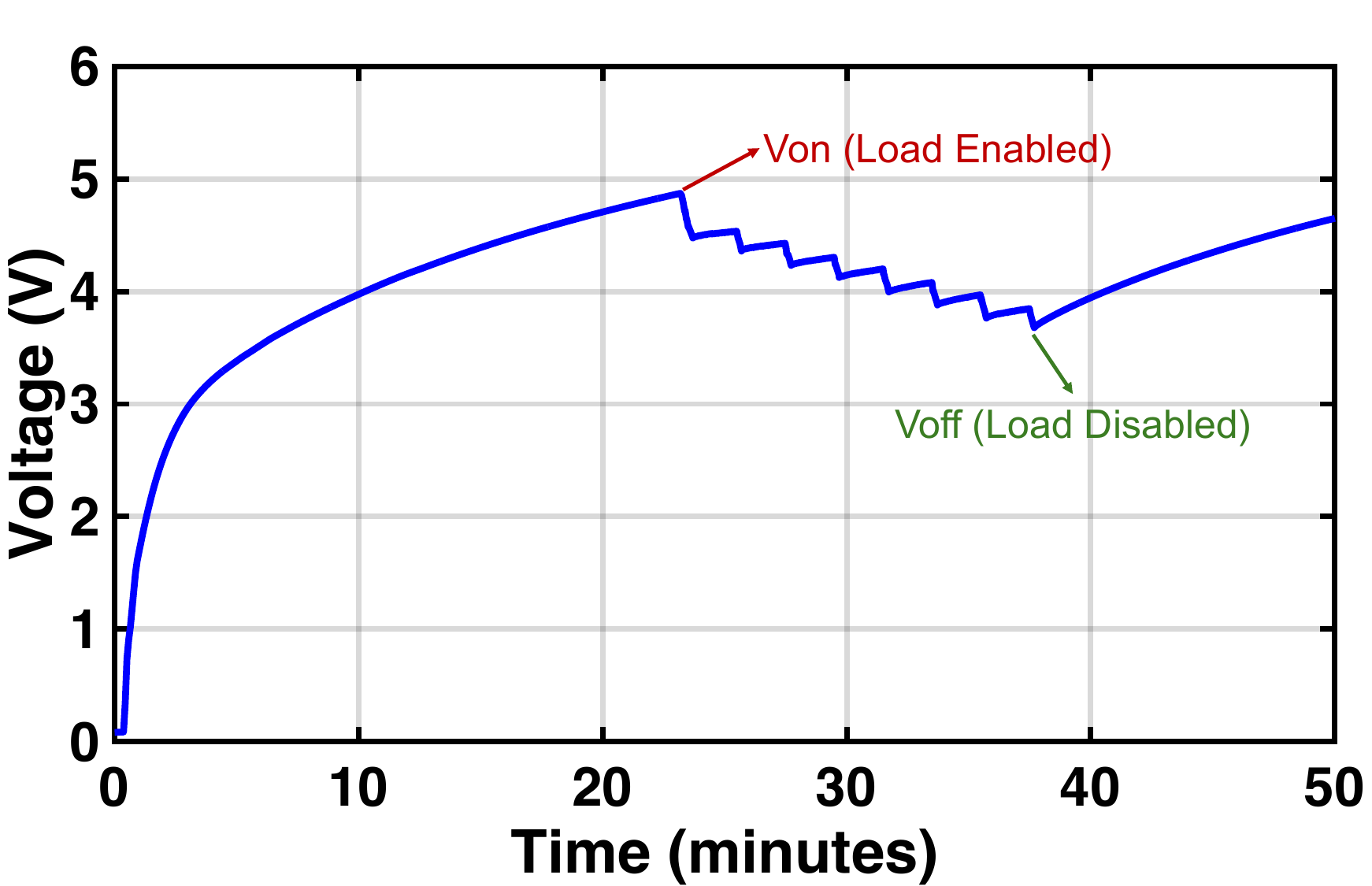}}
    \caption{Supercapacitor voltage profile showing comparator-based load control. The load activates at \(V_{\text{on}} = 4.87~\si{\volt}\) and deactivates at \(V_{\text{off}} = 3.67~\si{\volt}\), with periodic voltage drops corresponding to LTE-M beacon transmissions every 2 minutes.}

\label{fig:comparator_behavior}
\end{figure}

As shown in Fig.~\ref{fig:comparator_behavior}, the supercapacitor begins charging immediately after water exposure, powered by the sensor's electrochemical energy generation. It takes approximately 23 minutes for the capacitor to reach the turn-on threshold of $V_{\text{on}} = 4.87~\si{\volt}$, at which point the comparator circuit enables the MOSFET, connecting the LTE-M module to the power rail. The activation time was evaluated by conducting five trials for each sensor unit. The average activation time was 23 minutes, with a standard deviation of 4 minutes, indicating consistent power-up of the LTE-M module. While this activation latency is higher than that of continuously powered nodes, it offers two distinct operational advantages for infrastructure monitoring. First, the charging period effectively functions as a natural temporal filter, ensuring that alerts are triggered only by persistent liquid accumulation rather than transient splashes or cleaning events. Second, in the context of hidden structural leaks, which frequently remain undetected for weeks or months before causing mold or drywall damage, a response time of roughly 20 minutes represents an improvement over manual inspection cycles, providing timely intervention capability without the logistical burden of battery maintenance.

Following the initial transmission sequence, the system enters a continuous operational cycle. As shown in Fig.~\ref{fig:comparator_behavior}, the capacitor voltage rises during each idle phase, with periodic step-wise drops corresponding to successive beacon events. This charge–transmit–recharge sequence continues until the voltage falls to approximately 3.67~V, at which point the comparator deactivates the MOSFET, disconnecting the LTE-M module from the power rail and allowing the capacitor to resume uninterrupted charging. This behavior implements a hysteretic control strategy that guarantees activation only when sufficient stored energy is available. While ideal simulations estimated the turn-off threshold ($V_{\text{off}}$) to be approximately 3.25~V (see Fig.~\ref{fig:comparator_simulated}), real-world measurements revealed a higher cutoff voltage of 3.67~V—attributable to non-idealities such as resistor tolerances, comparator input bias currents, and MOSFET switching behavior. Although this higher cutoff voltage reduces the effective discharge window, the selection of a \SI{1.5}{\farad} supercapacitor provides a substantial energy margin. With a capacity roughly double the calculated requirement of \SI{0.71}{\farad} (Eq.~\ref{eq:cap_calc}), the system retains sufficient energy to complete the transmission cycle even with the elevated cutoff voltage.

To further balance energy availability with communication frequency, the LTE-M module is configured to wake at fixed 2-minute intervals to initiate beacon transmissions. This interval was selected as a practical baseline informed by preliminary empirical observations rather than formal optimization. Experimental trials demonstrated that a 2-minute idle window consistently allowed the 1.5~F supercapacitor to recharge sufficiently, preventing premature voltage collapse while maintaining a regular transmission cadence. In practice, this time-based interval also functions as a secondary gating mechanism that complements the TLV431 voltage thresholds, helping to suppress brownout conditions and maximize the number of successful transmissions per charge cycle. Although the 2-minute interval proved effective under the experimental conditions reported here, it remains fully configurable in firmware. Longer idle durations may be adopted in applications where maximizing transmission count per charge cycle is prioritized. Conversely, shorter intervals may be preferred in challenging RF environments (e.g., deep basements) to provide transmission redundancy, thereby mitigating the risk of packet loss caused by signal fading or interference without altering the hardware architecture.

Experimental results show that a single charge–discharge cycle of the 1.5~F supercapacitor reliably supports an average of eight consecutive beacon transmissions before the system transitions into an isolated state (see Fig.~\ref{fig:comparator_behavior}). Furthermore, because the system leverages the existing LTE-M cellular infrastructure, its communication range is inherently defined by the availability of network coverage, eliminating the need for additional gateway hardware or relay nodes.

\subsection{nRF9160 Power Consumption Analysis}
\label{subsec:power consumption analysis}

To evaluate the power requirements of the LTE-M communication subsystem, we analyzed the current consumption profile of the Nordic Thingy:91 development board, which integrates the nRF9160 SiP. The board was powered with a constant 4.87~V was used to emulate the fully charged state of the supercapacitor during activation. The current consumption was recorded using the Nordic Power Profiler Kit II at a sampling rate of 10,000 samples per second.

\begin{figure*}[!t]
    \centering
    \subfloat[Overall system behavior\label{fig:current_overall}]{
        \includegraphics[width=0.45\linewidth]{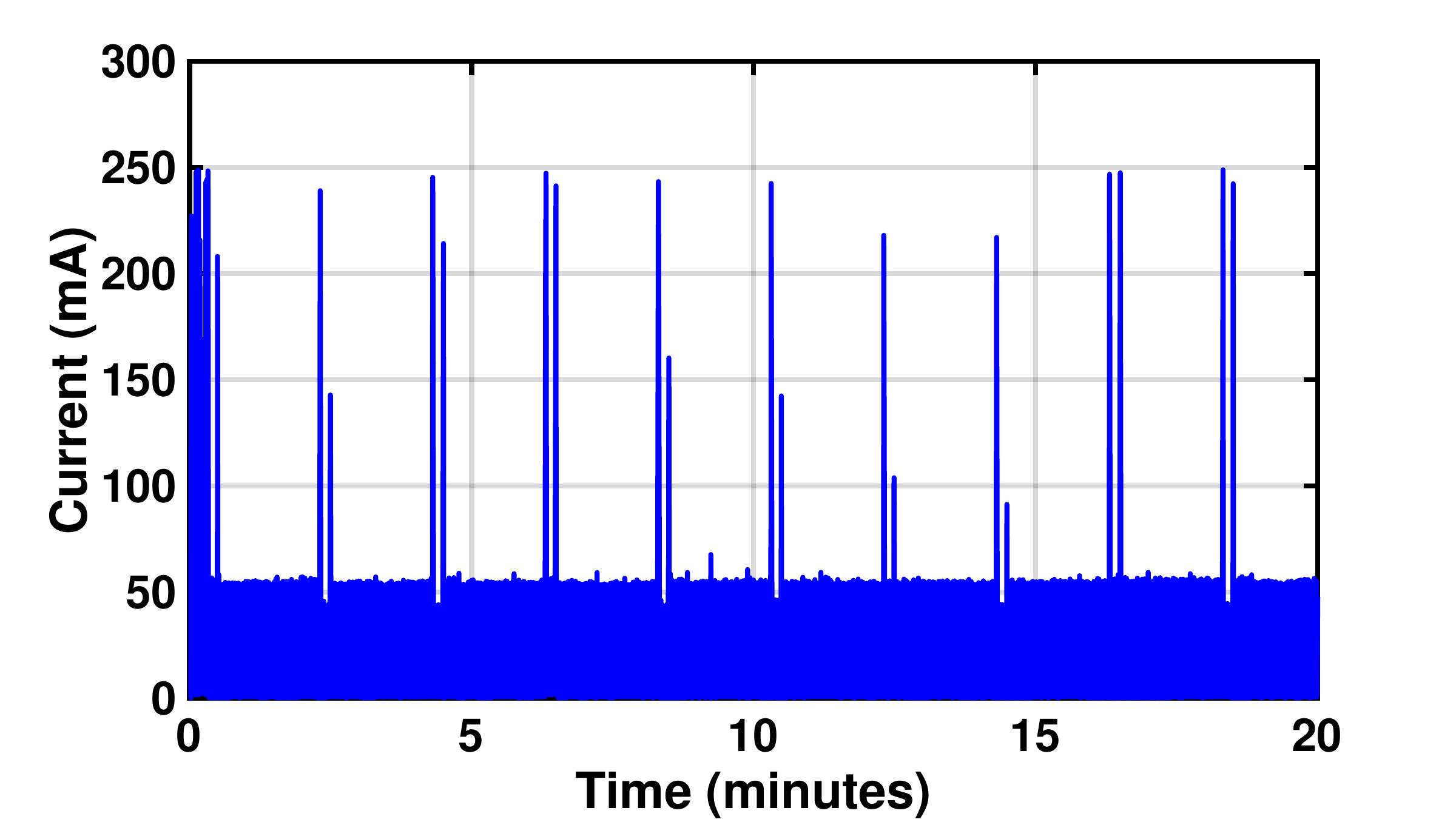}
    }
    \hfill
    \subfloat[Initial network search\label{fig:current_initial}]{
        \includegraphics[width=0.45\linewidth]{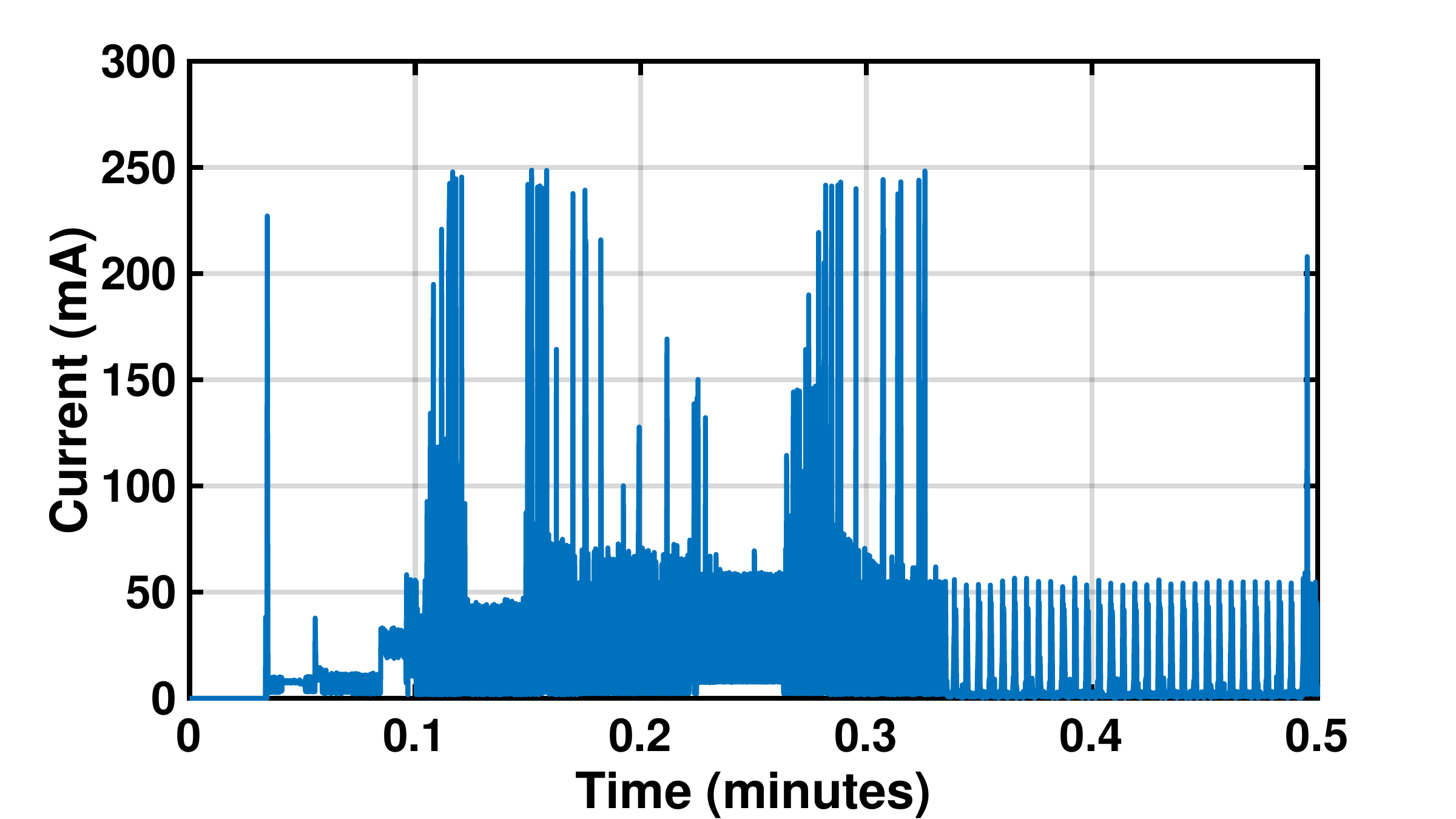}
    }
    \\
    \subfloat[Idle phase\label{fig:current_idle}]{
        \includegraphics[width=0.45\linewidth]{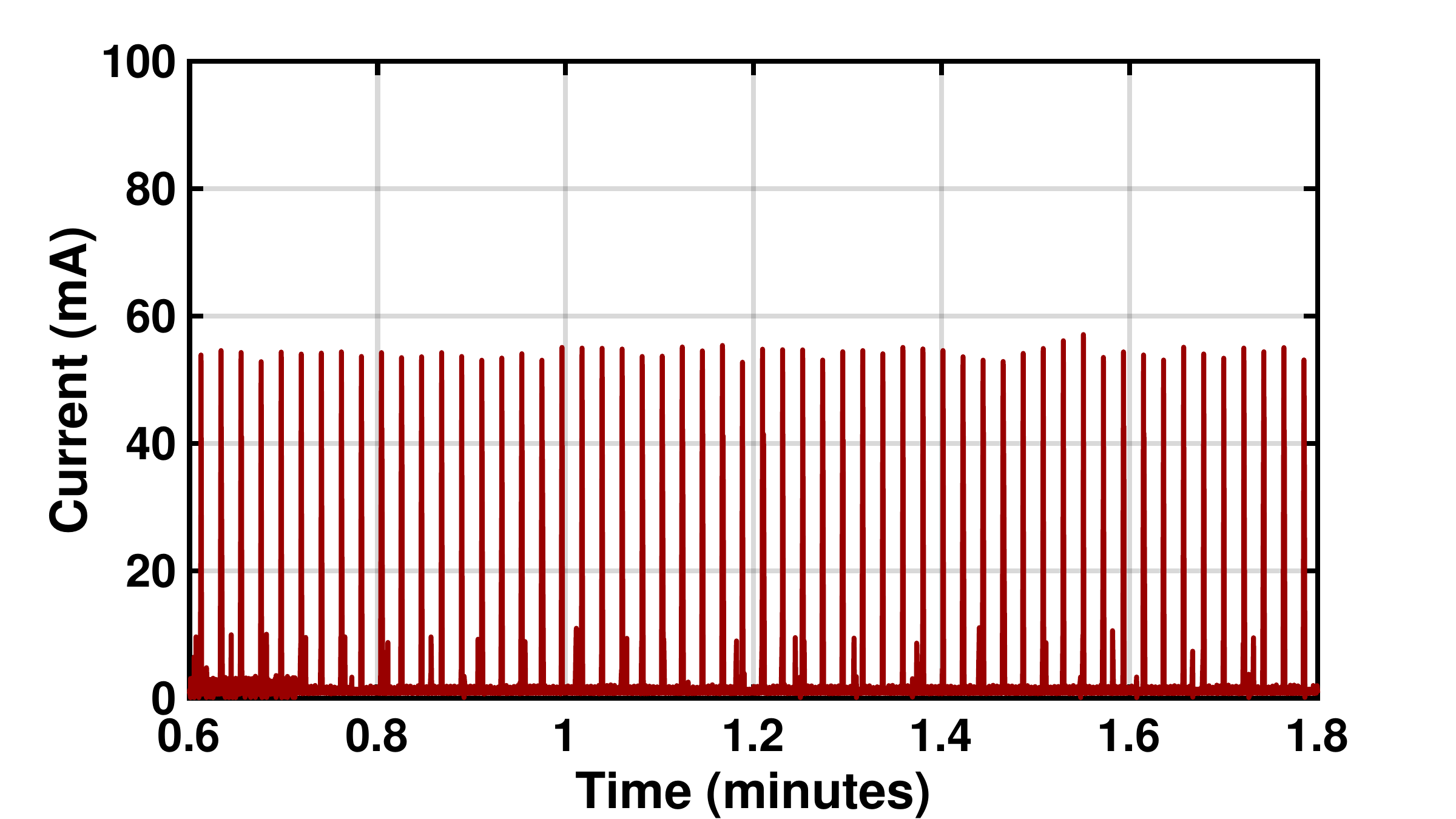}
    }
    \hfill
    \subfloat[Zoomed idle spikes\label{fig:current_idle_spikes}]{
        \includegraphics[width=0.45\linewidth]{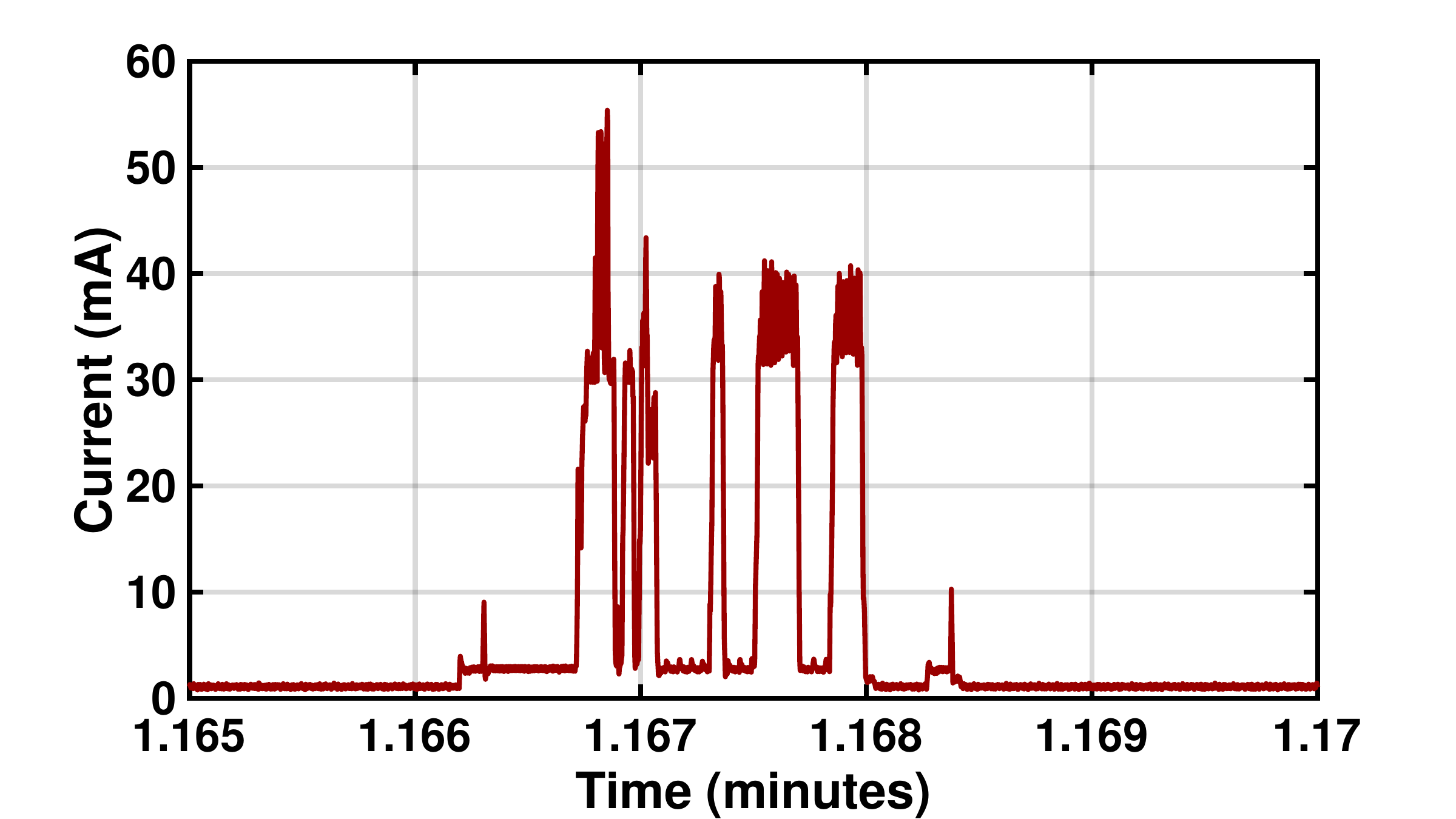}
    }
    \\
    \subfloat[Subsequent data transmission\label{fig:current_subsequent}]{
        \includegraphics[width=0.45\linewidth]{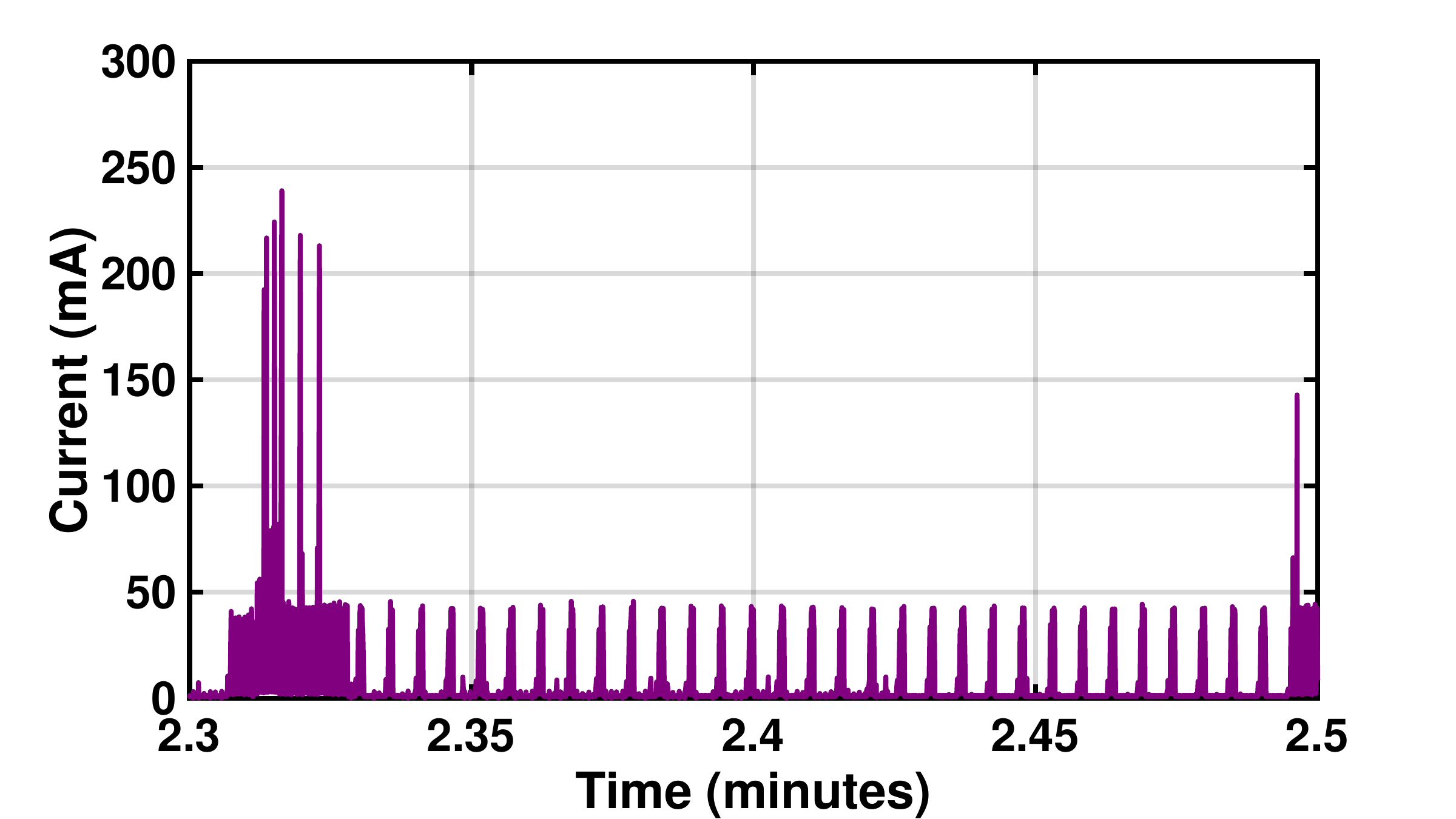}
    }

    \caption{Measured current consumption profile of the LTE-M system: 
    (a) overall 20-minute behavior showing full system operation; 
    (b) zoomed view of the initial network search phase; 
    (c) zoomed view of the idle phase between transmissions; 
    (d) detailed view showing brief current spikes during idle; 
    (e) zoomed view of subsequent periodic data transmissions.}
    \label{fig:current_profile}
\end{figure*}

Fig.~\ref{fig:current_profile} illustrates the current consumption behavior over time, including  initial network search, idle mode with PSM disabled, and periodic LTE-M uplink transmissions.

For each phase of operation, the average current and peak current were extracted from the recorded data. Energy consumption was computed by integrating the current-time profile using the trapezoidal rule, which estimates the area under the current-time curve to determine the total charge consumed. This value was then multiplied by the supply voltage to obtain energy in joules:
\begin{equation}
E = V \cdot \int_{t_1}^{t_2} I(t) \, dt \approx V \cdot \sum_{i} \frac{I_i + I_{i+1}}{2} \cdot \Delta t
\label{eq:energy_trapezoidal}
\end{equation}

For phases with relatively stable current (e.g., idle), energy was also cross-validated by multiplying average power with duration.

The initial network search phase spans approximately 30~s and corresponds to the LTE-M modem’s attachment and synchronization with the base station. This is the most energy-intensive stage, with a peak current of 248.74~mA and a total energy usage of 2.15~J.

During the idle state, which lasts 120~s, the current draw gets reduced. The peak and average current during this period were calculated at 57.2~mA and 1.72~mA respectively, resulting in an energy consumption of 0.98~J.

Finally, each uplink transmission phase, lasting approximately 12~s, includes brief bursts of high current during packet transmission. The peak current in this phase was 239.05~mA, and the total energy consumed was approximately 0.35~J.

The results are summarized in Table~\ref{tab:ltem_energy}, providing a breakdown of current, power, and energy for each operational phase of the LTE-M system.

\begin{table*}[t]
\centering
\caption{Measured Power and Energy Consumption During LTE-M Operation Phases}
\label{tab:ltem_energy}
\begin{tabular}{lccccc}
\toprule
Stage & Duration (s) & Peak Current (mA) & Avg. Current (mA) & Avg. Power (W) & Energy (J) \\
\midrule
Initial network search & 30 & 248.7 & 14.9 & 0.0717 & 2.15 \\
Idle & 120 & 57.2 & 1.72 & 0.0082 & 0.98 \\
Subsequent transmission & 12 & 239.1 & 6.12 & 0.0294 & 0.35 \\
\bottomrule
\end{tabular}
\end{table*}

\subsection{Sensitivity to Water Depth}

To assess the responsiveness of the energy harvesting system under different leak conditions, we evaluated its behavior at varying water depths: 0.5~mm, 1.0~mm, and 1.5~mm. Since the electrochemical reaction that drives energy generation is initiated by the presence of water, even small variations in depth can influence the ionic contact area and thereby affect the rate of energy accumulation. This makes water depth an important parameter for characterizing the sensor's practical sensitivity to real-world leak events.

\begin{figure}[!t]
\centerline{\includegraphics[width=80mm]{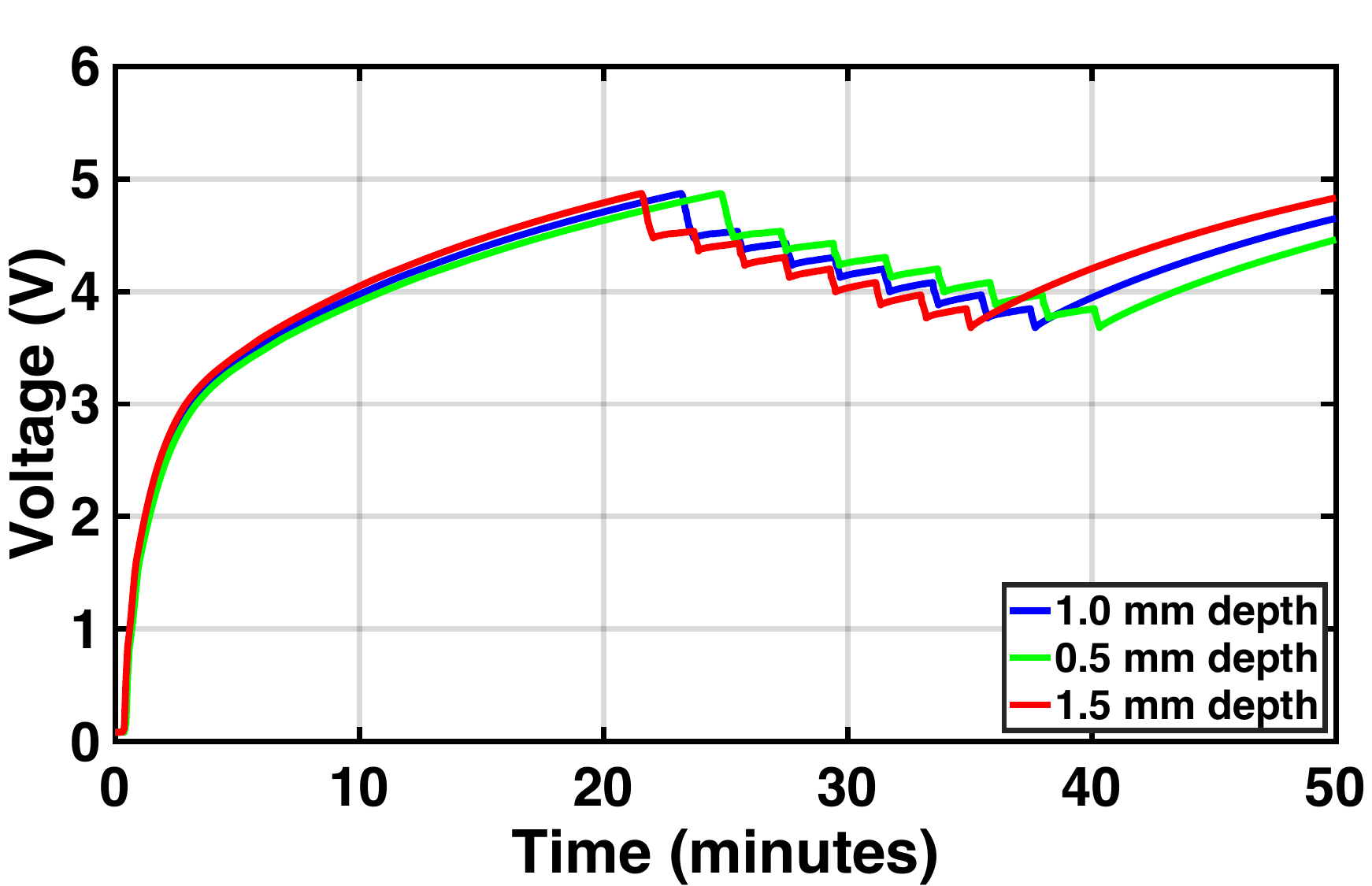}}
    \caption{Supercapacitor voltage profiles at varying water depths (0.5~mm, 1.0~mm, 1.5~mm), illustrating the sensitivity of the energy harvesting system to water level.}

\label{fig:sensitivity}
\end{figure}

In the experiment, three identical sensor units were individually tested by placing them in a Petri dish with controlled water depths of 0.5~mm, 1.0~mm, and 1.5~mm. The 1.5~F supercapacitor voltage was measured using a DMM to accurately monitor the charging behavior over time. The resulting voltage profiles are shown in Fig.~\ref{fig:sensitivity}. The time required for the capacitor to charge up to the activation threshold of approximately 4.87~V varies slightly with depth. At 1.5~mm, the system activates in around 21 minutes, while the 1.0~mm activates at 23 minutes. Even at the shallowest depth of 0.5~mm, the system successfully charges and activates within 24 minutes.

These results confirm that the system is capable of operating with very low amounts of water, down to 0.5~mm in depth. While the activation times differ only modestly, they effectively reflect the impact of water volume on electrochemical activity and energy harvesting rate. Furthermore, the electrochemical process is nonlinear and environmentally dependent. Factors such as water conductivity, electrode condition, and ambient temperature can cause slight variations in activation time across repeated trials.

It should be noted that the electrochemical generation rate is influenced by fluid conductivity. While the experimental results reported here utilized standard tap water, variations in water quality—such as high-salinity floodwater or deionized condensate—may impact the activation latency. Generally, higher ionic concentration accelerates the electrochemical reaction, potentially reducing the activation time, whereas highly purified water may extend the initial charge duration due to higher internal resistance. Regarding thermal stability, previous characterization of this sensor architecture has already demonstrated robust voltage generation across a wide temperature range (\SI{-20}{\degreeCelsius} to \SI{70}{\degreeCelsius})~\cite{Rouhi2024}. Therefore, regardless of specific water composition or environmental temperature, the electrochemical potential remains sufficient to initiate the reaction. As demonstrated by the system's high sensitivity to depths as low as \SI{0.5}{\milli\meter}, activation is ultimately guaranteed once the fluid bridges the contacts, ensuring consistent leak detection across diverse deployment scenarios.

\subsection{Boost Converter Efficiency}
\label{subsec:boost_converter_efficiency}

To evaluate the performance of the 5~V boost converter (ME2108A50) used in our system, we measured its power conversion efficiency using a Source Measure Unit (SMU) configured as a current sink. This setup allowed us to simulate different load conditions while accurately monitoring the converter’s input and output characteristics.

The converter was supplied with a fixed input voltage of 1.2~V, representative of the average voltage generated by the energy harvesting sensor under load. 
The output current was swept across a range of values using the SMU, while the input current was monitored via a separate DC source. Efficiency was computed using the conventional relation:

\begin{equation}
\eta = \frac{V_{\text{out}} \times I_{\text{out}}}{V_{\text{in}} \times I_{\text{in}}} \times 100\%
\label{eq:efficiency}
\end{equation}

\begin{figure}[!t]
\centerline{\includegraphics[width=80mm]{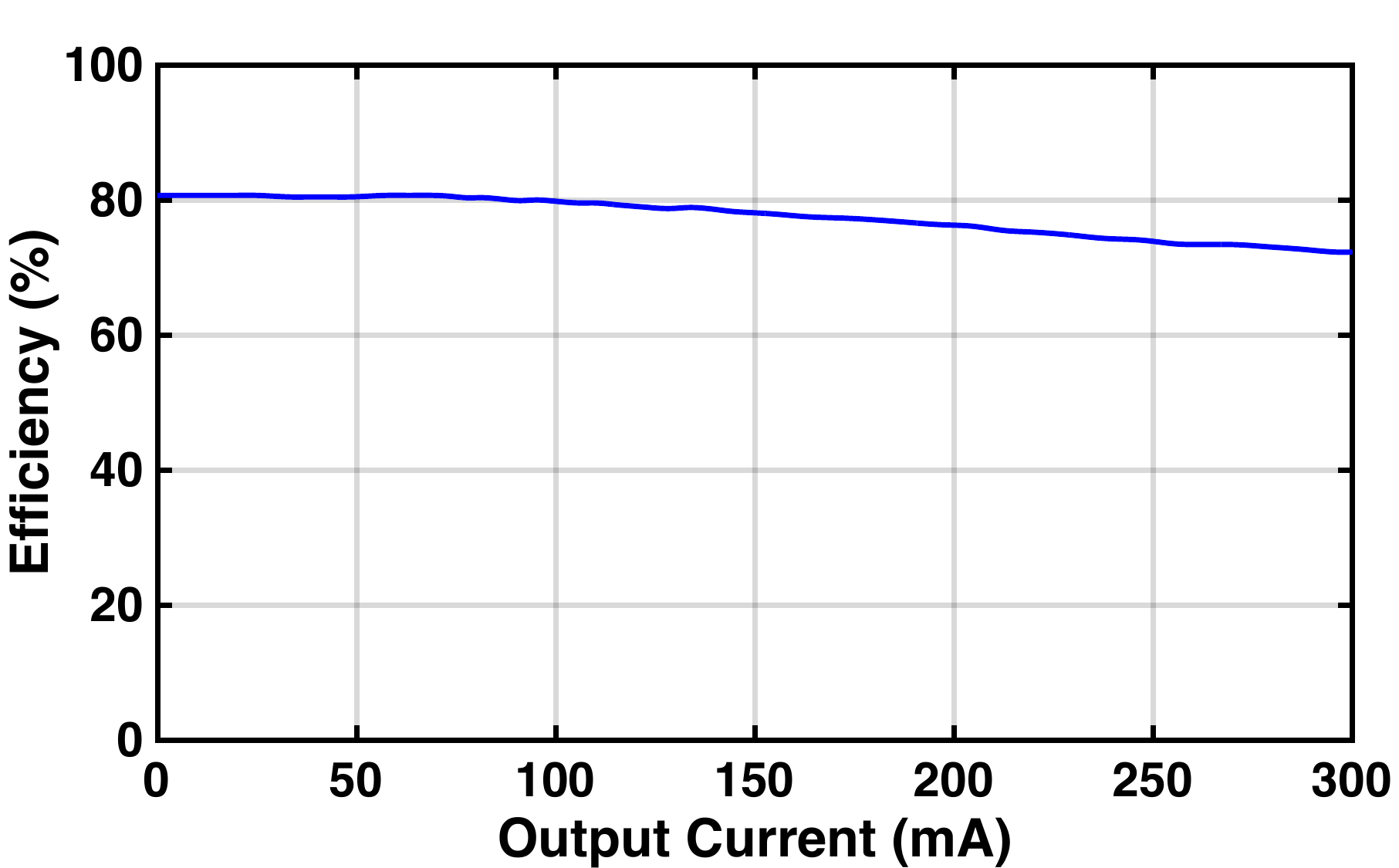}}
    \caption{Measured efficiency of the 5V boost converter at 1.2V input with varying output currents.}

\label{fig:boost_efficiency}
\end{figure}

The measured efficiency curve, shown in Fig.~\ref{fig:boost_efficiency}, indicates that the converter maintains an efficiency above 80\% under light load conditions, with a gradual decline as the output current increases. At an output current of 250~mA—corresponding to the peak demand observed during LTE-M module startup and data transmission, the converter achieves an efficiency of approximately 73\%. This performance confirms the converter’s capability to support the system’s most energy-intensive operations within acceptable efficiency margins.

While the ME2108A50 may not offer the highest efficiency compared to other commercial boost converters, such as the TPS61023 or MCP1642, it was retained for its superior low-voltage start-up capability as described in Section~\ref{subsec:boost_converter}. By contrast, most high-efficiency boost converters require a minimum input voltage of approximately 1.8~V. In our tests, this proved to be a limitation as the energy harvesting sensor, when loaded, often experienced voltage drops below 1.8~V, which prevented those converters from even turning on.

Thus, even though the ME2108A50 offers slightly lower peak efficiency, its ability to operate from very low voltages makes it well-suited for our application. The converter reliably regulates the output even when the sensor voltage falls significantly, ensuring uninterrupted energy storage and delivery for LTE-M communication in a battery-free, energy-constrained environment.

\subsection{Repeatability}
\label{subsec:repeatability}

The experimental results confirm that the system can sustain high-power LTE-M communication cycles using only harvested energy, without reliance on batteries or manual intervention. The consistent voltage profiles and predictable load-switching behavior across multiple trials demonstrate robust integration of sensing, energy conversion, power buffering, and wireless transmission in a fully battery-free configuration.

Following activation, the sensor continues to operate reliably until the water source is removed and the sensor begins to dry. As the hydration level decreases, electrochemical activity diminishes, leading to a decline in harvested energy. Although reintroducing water can trigger a limited response, the regenerated output is typically weak and inconsistent due to non-uniform ionic redistribution and corrosion of the electrode surface. Under the current design, the sensor therefore behaves effectively as a single-use component.

Reactivation would require refurbishment or replacement of the sensing element to restore consistent performance. However, this limitation is acceptable for most practical deployments, where leak events are infrequent and often trigger inspection or maintenance procedures. As such, the water sensor component can be conveniently replaced as part of the standard maintenance workflow without the need for replacing any of the underlying electronics. This modularity allows the system to maintain its battery-free, infrastructure-independent operation with minimal maintenance overhead.

\section{Conclusion}
\label{sec:conclusion}

This paper demonstrates a battery-free IoT water leak detection system that leverages LTE-M cellular communication without requiring local gateways. By combining a nanomaterial-based hydroelectric sensor utilizing abundant, non-toxic materials with efficient power management and a commercial LTE-M module, the system achieves direct-to-cloud communication triggered by the presence of water. Experimental results confirm reliable activation at water depths as low as 0.5~mm, sustained operation through energy-aware hardware gating, and the ability to meet the high power demands of LTE-M startup and transmission. The demonstrated satellite-compatible configuration further indicates the system’s potential for widespread IoT deployment, enabling connectivity in regions beyond terrestrial network coverage. As the first successful demonstration of this battery-free LTE-M architecture, the system serves as a baseline for future research. Subsequent iterations can further reduce activation latency and form factor through advancements in nanomaterial power density and the integration of ultra-low-power integrated circuits. Overall, the work establishes a practical foundation for low-maintenance, event-driven sensing platforms suited for infrastructure monitoring and other remote applications.

\section*{Acknowledgment}
The authors would like to acknowledge the support from MITACS, Rogers Communications, and AquaSensing Inc.

\balance
\bibliographystyle{IEEEtran}
\bibliography{references}

@misc{nordic_nrf9160,
  author       = {{Nordic Semiconductor}},
  title        = {{nRF9160 SiP Product Brief v2.1}},
  year         = {2023},
  url          = {https://www.nordicsemi.com/-/media/Software-and-other-downloads/Product-Briefs/nRF9160-SiP-PB-v2.1.pdf},
  note         = {Accessed: March 28, 2025}
}

@article{Michalski2021,
  title = {{Problems of Powering End Devices in Wireless Networks of the Internet of Things}},
  volume = {14},
  ISSN = {1996-1073},
  url = {http://dx.doi.org/10.3390/en14092417},
  DOI = {10.3390/en14092417},
  number = {9},
  journal = {Energies},
  publisher = {MDPI AG},
  author = {Michalski,  Andrzej and Watral,  Zbigniew},
  year = {2021},
  month = apr,
  pages = {2417}
}

@inproceedings{Nepal2024,
  title = {{Joint RF Energy Harvesting and Nanoelectronics for Self-Powered Water Leak Detection}},
  url = {http://dx.doi.org/10.23919/EuMC61614.2024.10732457},
  DOI = {10.23919/eumc61614.2024.10732457},
  booktitle = {2024 54th European Microwave Conference (EuMC)},
  publisher = {IEEE},
  author = {Nepal,  Roshan and Rouhi,  Mohammadreza and Zhou,  Norman and Shaker,  George},
  year = {2024},
  month = sep,
  pages = {876–879}
}

@article{nepal_Lora,
  title = {{Toward Long-Range,  Batteryless Water Leak Detection: A LoRa-Based Approach}},
  volume = {9},
  ISSN = {2475-1472},
  url = {http://dx.doi.org/10.1109/LSENS.2025.3609297},
  DOI = {10.1109/lsens.2025.3609297},
  number = {10},
  journal = {IEEE Sensors Letters},
  publisher = {Institute of Electrical and Electronics Engineers (IEEE)},
  author = {Nepal,  Roshan and Abbasi,  Roozbeh and Brown,  Brandon and Oginni,  Adunni and Zhou,  Norman and Shaker,  George},
  year = {2025},
  month = oct,
  pages = {1–4}
}

@article{Hasan2023,
  title = {{Navigating Battery Choices in IoT: An Extensive Survey of Technologies and Their Applications}},
  volume = {9},
  ISSN = {2313-0105},
  url = {http://dx.doi.org/10.3390/batteries9120580},
  DOI = {10.3390/batteries9120580},
  number = {12},
  journal = {Batteries},
  publisher = {MDPI AG},
  author = {Hasan,  Kareeb and Tom,  Neil and Yuce,  Mehmet Rasit},
  year = {2023},
  month = dec,
  pages = {580}
}

@article{Nikoukar2018,
  title = {{Low-Power Wireless for the Internet of Things: Standards and Applications}},
  volume = {6},
  ISSN = {2169-3536},
  url = {http://dx.doi.org/10.1109/ACCESS.2018.2879189},
  DOI = {10.1109/access.2018.2879189},
  journal = {IEEE Access},
  publisher = {Institute of Electrical and Electronics Engineers (IEEE)},
  author = {Nikoukar,  Ali and Raza,  Saleem and Poole,  Angelina and Gunes,  Mesut and Dezfouli,  Behnam},
  year = {2018},
  pages = {67893–67926}
}

@article{Jamshed2022,
  title = {{Challenges,  Applications,  and Future of Wireless Sensors in Internet of Things: A Review}},
  volume = {22},
  ISSN = {2379-9153},
  url = {http://dx.doi.org/10.1109/JSEN.2022.3148128},
  DOI = {10.1109/jsen.2022.3148128},
  number = {6},
  journal = {IEEE Sensors Journal},
  publisher = {Institute of Electrical and Electronics Engineers (IEEE)},
  author = {Jamshed,  Muhammad Ali and Ali,  Kamran and Abbasi,  Qammer H. and Imran,  Muhammad Ali and Ur-Rehman,  Masood},
  year = {2022},
  month = mar,
  pages = {5482–5494}
}

@article{Koulouras2025,
  title = {{Evolution of Bluetooth Technology: BLE in the IoT Ecosystem}},
  volume = {25},
  ISSN = {1424-8220},
  url = {http://dx.doi.org/10.3390/s25040996},
  DOI = {10.3390/s25040996},
  number = {4},
  journal = {Sensors},
  publisher = {MDPI AG},
  author = {Koulouras,  Grigorios and Katsoulis,  Stylianos and Zantalis,  Fotios},
  year = {2025},
  month = feb,
  pages = {996}
}

@article{Rouhi2024,
  title = {{Wireless Battery-Free Self-Powered Water Leak Detection Through Hydroelectric Energy Harvesting}},
  volume = {24},
  ISSN = {2379-9153},
  url = {http://dx.doi.org/10.1109/JSEN.2024.3469632},
  DOI = {10.1109/jsen.2024.3469632},
  number = {22},
  journal = {IEEE Sensors Journal},
  publisher = {Institute of Electrical and Electronics Engineers (IEEE)},
  author = {Rouhi,  Mohammadreza and Nepal,  Roshan and Chathanat,  Simran and Kotak,  Nimesh and Johnston,  Nathan and Ansariyan,  Ahmad and Kaur,  Kamalpreet and Patel,  Kushant and Zhou,  Norman and Shaker,  George},
  year = {2024},
  month = nov,
  pages = {37822–37835}
}

@article{Jeon2018,
  title = {{BLE Beacons for Internet of Things Applications: Survey,  Challenges,  and Opportunities}},
  volume = {5},
  ISSN = {2327-4662},
  url = {http://dx.doi.org/10.1109/JIOT.2017.2788449},
  DOI = {10.1109/jiot.2017.2788449},
  number = {2},
  journal = {IEEE Internet of Things Journal},
  publisher = {Institute of Electrical and Electronics Engineers (IEEE)},
  author = {Jeon,  Kang Eun and She,  James and Soonsawad,  Perm and Ng,  Pai Chet},
  year = {2018},
  month = apr,
  pages = {811–828}
}

@article{Naifar2024,
  title = {{Energy Harvesting Technologies and Applications for the Internet of Things and Wireless Sensor Networks}},
  volume = {24},
  ISSN = {1424-8220},
  url = {http://dx.doi.org/10.3390/s24144688},
  DOI = {10.3390/s24144688},
  number = {14},
  journal = {Sensors},
  publisher = {MDPI AG},
  author = {Naifar,  Slim and Kanoun,  Olfa and Trigona,  Carlo},
  year = {2024},
  month = jul,
  pages = {4688}
}

@article{Zohourian2023,
  title = {{IoT Zigbee device security: A comprehensive review}},
  volume = {22},
  ISSN = {2542-6605},
  url = {http://dx.doi.org/10.1016/j.iot.2023.100791},
  DOI = {10.1016/j.iot.2023.100791},
  journal = {Internet of Things},
  publisher = {Elsevier BV},
  author = {Zohourian,  Alireza and Dadkhah,  Sajjad and Neto,  Euclides Carlos Pinto and Mahdikhani,  Hassan and Danso,  Priscilla Kyei and Molyneaux,  Heather and Ghorbani,  Ali A.},
  year = {2023},
  month = jul,
  pages = {100791}
}

@article{AlShuhail2022,
  title = {{Zigbee-Based Low Power Consumption Wearables Device for Voice Data Transmission}},
  volume = {14},
  ISSN = {2071-1050},
  url = {http://dx.doi.org/10.3390/su141710847},
  DOI = {10.3390/su141710847},
  number = {17},
  journal = {Sustainability},
  publisher = {MDPI AG},
  author = {AlShuhail,  Asma Shuhail and Bhatia,  Surbhi and Kumar,  Ankit and Bhushan,  Bharat},
  year = {2022},
  month = aug,
  pages = {10847}
}

@article{Chilamkurthy2022,
  title = {{Low-Power Wide-Area Networks: A Broad Overview of Its Different Aspects}},
  volume = {10},
  ISSN = {2169-3536},
  url = {http://dx.doi.org/10.1109/ACCESS.2022.3196182},
  DOI = {10.1109/access.2022.3196182},
  journal = {IEEE Access},
  publisher = {Institute of Electrical and Electronics Engineers (IEEE)},
  author = {Chilamkurthy,  Naga Srinivasarao and Pandey,  Om Jee and Ghosh,  Anirban and Cenkeramaddi,  Linga Reddy and Dai,  Hong-Ning},
  year = {2022},
  pages = {81926–81959}
}

@article{Delgado2021,
  title = {{Batteryless LoRaWAN Communications Using Energy Harvesting: Modeling and Characterization}},
  volume = {8},
  ISSN = {2372-2541},
  url = {http://dx.doi.org/10.1109/JIOT.2020.3019140},
  DOI = {10.1109/jiot.2020.3019140},
  number = {4},
  journal = {IEEE Internet of Things Journal},
  publisher = {Institute of Electrical and Electronics Engineers (IEEE)},
  author = {Delgado,  Carmen and Sanz,  Jose Maria and Blondia,  Chris and Famaey,  Jeroen},
  year = {2021},
  month = feb,
  pages = {2694–2711}
}

@article{Toro2022,
  title = {{Backscatter Wireless Communications and Sensing in Green Internet of Things}},
  volume = {6},
  ISSN = {2473-2400},
  url = {http://dx.doi.org/10.1109/TGCN.2021.3095792},
  DOI = {10.1109/tgcn.2021.3095792},
  number = {1},
  journal = {IEEE Transactions on Green Communications and Networking},
  publisher = {Institute of Electrical and Electronics Engineers (IEEE)},
  author = {Toro,  Usman Saleh and Wu,  Kaishun and Leung,  Victor C. M.},
  year = {2022},
  month = mar,
  pages = {37–55}
}

@book{Gong2024,
  title = {{Practical Backscatter Communication for the Internet of Things}},
  ISBN = {9783031592546},
  ISSN = {2191-5776},
  url = {http://dx.doi.org/10.1007/978-3-031-59254-6},
  DOI = {10.1007/978-3-031-59254-6},
  journal = {SpringerBriefs in Computer Science},
  publisher = {Springer Nature Switzerland},
  author = {Gong,  Wei and Liu,  Jiangchuan and Wu,  Weiqi},
  year = {2024}
}

@misc{enocean2020whitepaper,
  title        = {EnOcean – The World of Energy Harvesting Wireless Technology},
  author       = {{EnOcean GmbH}},
  year         = {2020},
  month        = {February},
  url          = {https://www.enocean.com/wp-content/uploads/redaktion/pdf/white_paper/WhitePaper_Getting_Started_With_EnOcean_v3.0.pdf},
  note         = {Accessed: 2025-03-25}
}

@article{Moges2023,
  title = {{Cellular Internet of Things: Use cases,  technologies,  and future work}},
  volume = {24},
  ISSN = {2542-6605},
  url = {http://dx.doi.org/10.1016/j.iot.2023.100910},
  DOI = {10.1016/j.iot.2023.100910},
  journal = {Internet of Things},
  publisher = {Elsevier BV},
  author = {Moges,  Teshager Hailemariam and Lakew,  Demeke Shumeye and Nguyen,  Ngoc Phi and Dao,  Nhu-Ngoc and Cho,  Sungrae},
  year = {2023},
  month = dec,
  pages = {100910}
}

@article{Wang2021,
  title = {{Location-Based Timing Advance Estimation for 5G Integrated LEO Satellite Communications}},
  volume = {70},
  ISSN = {1939-9359},
  url = {http://dx.doi.org/10.1109/TVT.2021.3079936},
  DOI = {10.1109/tvt.2021.3079936},
  number = {6},
  journal = {IEEE Transactions on Vehicular Technology},
  publisher = {Institute of Electrical and Electronics Engineers (IEEE)},
  author = {Wang,  Wenjin and Chen,  Tingting and Ding,  Rui and Seco-Granados,  Gonzalo and You,  Li and Gao,  Xiqi},
  year = {2021},
  month = jun,
  pages = {6002–6017}
}

@article{Vaezi2022,
  title = {{Cellular,  Wide-Area,  and Non-Terrestrial IoT: A Survey on 5G Advances and the Road Toward 6G}},
  volume = {24},
  ISSN = {2373-745X},
  url = {http://dx.doi.org/10.1109/COMST.2022.3151028},
  DOI = {10.1109/comst.2022.3151028},
  number = {2},
  journal = {IEEE Communications Surveys \& Tutorials},
  publisher = {Institute of Electrical and Electronics Engineers (IEEE)},
  author = {Vaezi,  Mojtaba and Azari,  Amin and Khosravirad,  Saeed R. and Shirvanimoghaddam,  Mahyar and Azari,  M. Mahdi and Chasaki,  Danai and Popovski,  Petar},
  year = {2022},
  pages = {1117–1174}
}

@article{Choudhary2024,
  title = {{Internet of Things: A comprehensive overview,  architectures,  applications,  simulation tools,  challenges and future directions}},
  volume = {4},
  ISSN = {2730-7239},
  url = {http://dx.doi.org/10.1007/s43926-024-00084-3},
  DOI = {10.1007/s43926-024-00084-3},
  number = {1},
  journal = {Discover Internet of Things},
  publisher = {Springer Science and Business Media LLC},
  author = {Choudhary,  Anita},
  year = {2024},
  month = dec 
}

@article{Saad2024,
  title = {{Non-Terrestrial Networks: An Overview of 3GPP Release 17 \& 18}},
  volume = {7},
  ISSN = {2576-3199},
  url = {http://dx.doi.org/10.1109/IOTM.001.2300154},
  DOI = {10.1109/iotm.001.2300154},
  number = {1},
  journal = {IEEE Internet of Things Magazine},
  publisher = {Institute of Electrical and Electronics Engineers (IEEE)},
  author = {Saad,  Malik Muhammad and Tariq,  Muhammad Ashar and Khan,  Muhammad Toaha Raza and Kim,  Dongkyun},
  year = {2024},
  month = jan,
  pages = {20–26}
}

@article{Centenaro2021,
  title = {{A Survey on Technologies,  Standards and Open Challenges in Satellite IoT}},
  volume = {23},
  ISSN = {2373-745X},
  url = {http://dx.doi.org/10.1109/COMST.2021.3078433},
  DOI = {10.1109/comst.2021.3078433},
  number = {3},
  journal = {IEEE Communications Surveys \& Tutorials},
  publisher = {Institute of Electrical and Electronics Engineers (IEEE)},
  author = {Centenaro,  Marco and Costa,  Cristina E. and Granelli,  Fabrizio and Sacchi,  Claudio and Vangelista,  Lorenzo},
  year = {2021},
  pages = {1693–1720}
}

@inbook{Raghunandan2022,
  title = {{Satellite Communication}},
  ISBN = {9783030921880},
  ISSN = {2524-4353},
  url = {http://dx.doi.org/10.1007/978-3-030-92188-0_13},
  DOI = {10.1007/978-3-030-92188-0_13},
  booktitle = {Introduction to Wireless Communications and Networks},
  publisher = {Springer International Publishing},
  author = {Raghunandan,  Krishnamurthy},
  year = {2022},
  pages = {247–275}
}

@article{Qu2017,
  title = {{LEO Satellite Constellation for Internet of Things}},
  volume = {5},
  ISSN = {2169-3536},
  url = {http://dx.doi.org/10.1109/ACCESS.2017.2735988},
  DOI = {10.1109/access.2017.2735988},
  journal = {IEEE Access},
  publisher = {Institute of Electrical and Electronics Engineers (IEEE)},
  author = {Qu,  Zhicheng and Zhang,  Gengxin and Cao,  Haotong and Xie,  Jidong},
  year = {2017},
  pages = {18391–18401}
}

@article{Osoro2021,
  title = {{A Techno-Economic Framework for Satellite Networks Applied to Low Earth Orbit Constellations: Assessing Starlink,  OneWeb and Kuiper}},
  volume = {9},
  ISSN = {2169-3536},
  url = {http://dx.doi.org/10.1109/ACCESS.2021.3119634},
  DOI = {10.1109/access.2021.3119634},
  journal = {IEEE Access},
  publisher = {Institute of Electrical and Electronics Engineers (IEEE)},
  author = {Osoro,  Ogutu B. and Oughton,  Edward J.},
  year = {2021},
  pages = {141611–141625}
}

@article{elikbilek2022,
  title = {{Survey on Optimization Methods for LEO-Satellite-Based Networks with Applications in Future Autonomous Transportation}},
  volume = {22},
  ISSN = {1424-8220},
  url = {http://dx.doi.org/10.3390/s22041421},
  DOI = {10.3390/s22041421},
  number = {4},
  journal = {Sensors},
  publisher = {MDPI AG},
  author = {undefinedelikbilek,  Kaan and Saleem,  Zainab and Morales Ferre,  Ruben and Praks,  Jaan and Lohan,  Elena Simona},
  year = {2022},
  month = feb,
  pages = {1421}
}

@article{Hu2021,
  title = {{Self-powered 5G NB-IoT system for remote monitoring applications}},
  volume = {87},
  ISSN = {2211-2855},
  url = {http://dx.doi.org/10.1016/j.nanoen.2021.106140},
  DOI = {10.1016/j.nanoen.2021.106140},
  journal = {Nano Energy},
  publisher = {Elsevier BV},
  author = {Hu,  Guosheng and Yi,  Zhiran and Lu,  Lijun and Huang,  Yang and Zhai,  Yueqi and Liu,  Jingquan and Yang,  Bin},
  year = {2021},
  month = sep,
  pages = {106140}
}

@article{Yang2021,
  title = {{A First Look at Energy Consumption of NB-IoT in the Wild: Tools and Large-Scale Measurement}},
  volume = {29},
  ISSN = {1558-2566},
  url = {http://dx.doi.org/10.1109/TNET.2021.3096656},
  DOI = {10.1109/tnet.2021.3096656},
  number = {6},
  journal = {IEEE/ACM Transactions on Networking},
  publisher = {Institute of Electrical and Electronics Engineers (IEEE)},
  author = {Yang,  Deliang and Huang,  Xuan and Huang,  Jun and Chang,  Xiangmao and Xing,  Guoliang and Yang,  Yang},
  year = {2021},
  month = dec,
  pages = {2616–2631}
}

@article{Sultania2023,
  title = {{Batteryless NB-IoT prototype for bidirectional communication powered by ambient light}},
  volume = {142},
  ISSN = {1570-8705},
  url = {http://dx.doi.org/10.1016/j.adhoc.2023.103100},
  DOI = {10.1016/j.adhoc.2023.103100},
  journal = {Ad Hoc Networks},
  publisher = {Elsevier BV},
  author = {Sultania,  Ashish Kumar and Famaey,  Jeroen},
  year = {2023},
  month = apr,
  pages = {103100}
}

@inproceedings{AragonesOrtiz2020,
  series = {20ADIP},
  title = {{The Green Revolution for Oil\&Gas Using Battery-Less NB-IoT IIoT Devices Powered by Waste Heat for Process Maintenance}},
  url = {http://dx.doi.org/10.2118/203066-MS},
  DOI = {10.2118/203066-ms},
  booktitle = {Day 3 Wed,  November 11,  2020},
  publisher = {SPE},
  author = {Aragones Ortiz,  Raul and Nicolas Alegret,  Roger and Oliver Malagelada,  Joan and Malet Munté,  Roger and Ferrer Ramis,  Carles and Comellas Vogel,  David and Voces Merayo,  Ramon},
  year = {2020},
  month = nov,
  collection = {20ADIP}
}

@article{Liu2021,
  title = {{RACH in Self-Powered NB-IoT Networks: Energy Availability and Performance Evaluation}},
  volume = {69},
  ISSN = {1558-0857},
  url = {http://dx.doi.org/10.1109/TCOMM.2020.3041751},
  DOI = {10.1109/tcomm.2020.3041751},
  number = {3},
  journal = {IEEE Transactions on Communications},
  publisher = {Institute of Electrical and Electronics Engineers (IEEE)},
  author = {Liu,  Yan and Deng,  Yansha and Elkashlan,  Maged and Nallanathan,  Arumugam and Yuan,  Jinhong and Mallik,  Ranjan K.},
  year = {2021},
  month = mar,
  pages = {1750–1764}
}

@inbook{Borkar2020,
  title = {{Long-term evolution for machines (LTE-M)}},
  ISBN = {9780128188804},
  url = {http://dx.doi.org/10.1016/B978-0-12-818880-4.00007-7},
  DOI = {10.1016/b978-0-12-818880-4.00007-7},
  booktitle = {LPWAN Technologies for IoT and M2M Applications},
  publisher = {Elsevier},
  author = {Borkar,  Suresh R.},
  year = {2020},
  pages = {145–166}
}

@article{Bhilwaria2021,
  title = {{Critical power analysis for control path of a CAT-M based edge device}},
  volume = {13},
  ISSN = {2511-2112},
  url = {http://dx.doi.org/10.1007/s41870-021-00640-y},
  DOI = {10.1007/s41870-021-00640-y},
  number = {3},
  journal = {International Journal of Information Technology},
  publisher = {Springer Science and Business Media LLC},
  author = {Bhilwaria,  Harshita and Ranga,  Virender and Gargi,  Anirudh},
  year = {2021},
  month = apr,
  pages = {845–855}
}

@inproceedings{Labdaoui2023,
  title = {{Energy-efficient IoT Communications: A Comparative Study of Long-Term Evolution for Machines (LTE-M) and Narrowband Internet of Things (NB-IoT) Technologies}},
  url = {http://dx.doi.org/10.1109/ISCC58397.2023.10218061},
  DOI = {10.1109/iscc58397.2023.10218061},
  booktitle = {2023 IEEE Symposium on Computers and Communications (ISCC)},
  publisher = {IEEE},
  author = {Labdaoui,  Nassim and Nouvel,  Fabienne and Dutertre,  Stéphane},
  year = {2023},
  month = jul 
}

@misc{nordic2022thingy91,
  author       = {{Nordic Semiconductor}},
  title        = {{Nordic Thingy:91 – Multi-sensor Prototyping Platform for Cellular IoT and GNSS, Product Brief v2.1}},
  year         = {2022},
  howpublished = {\url{https://www.nordicsemi.com/-/media/Software-and-other-downloads/Product-Briefs/Nordic-Thingy-91-PB-v2.1.pdf}},
  note         = {Accessed: Mar. 26, 2025}
}

@article{Aragones2022,
  title = {{Autonomous Battery-Less Vibration IIoT Powered by Waste Heat for Chemical Plants Using NB-IoT}},
  volume = {22},
  ISSN = {2379-9153},
  url = {http://dx.doi.org/10.1109/JSEN.2022.3184267},
  DOI = {10.1109/jsen.2022.3184267},
  number = {15},
  journal = {IEEE Sensors Journal},
  publisher = {Institute of Electrical and Electronics Engineers (IEEE)},
  author = {Aragones,  Raul and Alegret,  Roger Nicolas and Oliver,  Joan and Ferrer,  Carles},
  year = {2022},
  month = aug,
  pages = {15448–15456}
}

@article{Islam2022,
  title = {{A Review on Current Technologies and Future Direction of Water Leakage Detection in Water Distribution Network}},
  volume = {10},
  ISSN = {2169-3536},
  url = {http://dx.doi.org/10.1109/ACCESS.2022.3212769},
  DOI = {10.1109/access.2022.3212769},
  journal = {IEEE Access},
  publisher = {Institute of Electrical and Electronics Engineers (IEEE)},
  author = {Islam,  Mohammed Rezwanul and Azam,  Sami and Shanmugam,  Bharanidharan and Mathur,  Deepika},
  year = {2022},
  pages = {107177–107201}
}

@article{Jan2022,
  title = {{IoT-Based Solutions to Monitor Water Level,  Leakage,  and Motor Control for Smart Water Tanks}},
  volume = {14},
  ISSN = {2073-4441},
  url = {http://dx.doi.org/10.3390/w14030309},
  DOI = {10.3390/w14030309},
  number = {3},
  journal = {Water},
  publisher = {MDPI AG},
  author = {Jan,  Farmanullah and Min-Allah,  Nasro and Saeed,  Saqib and Iqbal,  Sardar Zafar and Ahmed,  Rashad},
  year = {2022},
  month = jan,
  pages = {309}
}

@article{Bakhtawar2023,
  title = {{State‐of‐the‐art review of leak diagnostic experiments: Toward a smart water network}},
  volume = {10},
  ISSN = {2049-1948},
  url = {http://dx.doi.org/10.1002/wat2.1667},
  DOI = {10.1002/wat2.1667},
  number = {5},
  journal = {WIREs Water},
  publisher = {Wiley},
  author = {Bakhtawar,  Beenish and Zayed,  Tarek},
  year = {2023},
  month = may 
}

@article{Alshami2024,
  title = {{IoT Innovations in Sustainable Water and Wastewater Management and Water Quality Monitoring: A Comprehensive Review of Advancements,  Implications,  and Future Directions}},
  volume = {12},
  ISSN = {2169-3536},
  url = {http://dx.doi.org/10.1109/ACCESS.2024.3392573},
  DOI = {10.1109/access.2024.3392573},
  journal = {IEEE Access},
  publisher = {Institute of Electrical and Electronics Engineers (IEEE)},
  author = {Alshami,  Ahmad and Ali,  Eslam and Elsayed,  Moustafa and Eltoukhy,  Abdelrahman E. E. and Zayed,  Tarek},
  year = {2024},
  pages = {58427–58453}
}

@article{Cao2024,
  title = {{Wireless Battery Management Systems: Innovations,  Challenges,  and Future Perspectives}},
  volume = {17},
  ISSN = {1996-1073},
  url = {http://dx.doi.org/10.3390/en17133277},
  DOI = {10.3390/en17133277},
  number = {13},
  journal = {Energies},
  publisher = {MDPI AG},
  author = {Cao,  Zhi and Gao,  Wei and Fu,  Yuhong and Mi,  Chris},
  year = {2024},
  month = jul,
  pages = {3277}
}

@inproceedings{Zachariah2015,
  series = {HotMobile ’15},
  title = {{The Internet of Things Has a Gateway Problem}},
  url = {http://dx.doi.org/10.1145/2699343.2699344},
  DOI = {10.1145/2699343.2699344},
  booktitle = {Proceedings of the 16th International Workshop on Mobile Computing Systems and Applications},
  publisher = {ACM},
  author = {Zachariah,  Thomas and Klugman,  Noah and Campbell,  Bradford and Adkins,  Joshua and Jackson,  Neal and Dutta,  Prabal},
  year = {2015},
  month = feb,
  pages = {27–32},
  collection = {HotMobile ’15}
}

@article{AkinPonnle2021,
  title = {{Energy Harvesting Mechanisms in a Smart City—A Review}},
  volume = {4},
  ISSN = {2624-6511},
  url = {http://dx.doi.org/10.3390/smartcities4020025},
  DOI = {10.3390/smartcities4020025},
  number = {2},
  journal = {Smart Cities},
  publisher = {MDPI AG},
  author = {Akin-Ponnle,  Ajibike Eunice and Carvalho,  Nuno Borges},
  year = {2021},
  month = apr,
  pages = {476–498}
}

@article{Safaei2025,
  title = {{Eco-Friendly IoT: Leveraging Energy Harvesting for a Sustainable Future}},
  ISSN = {2995-7478},
  url = {http://dx.doi.org/10.1109/SR.2025.3552043},
  DOI = {10.1109/sr.2025.3552043},
  journal = {IEEE Sensors Reviews},
  publisher = {Institute of Electrical and Electronics Engineers (IEEE)},
  author = {Safaei,  Bardia and Peiravian,  Mana and Siamaki,  Mahdi},
  year = {2025},
  pages = {1–44}
}

@article{Pereira2020,
  title = {{Challenges in Resource-Constrained IoT Devices: Energy and Communication as Critical Success Factors for Future IoT Deployment}},
  volume = {20},
  ISSN = {1424-8220},
  url = {http://dx.doi.org/10.3390/s20226420},
  DOI = {10.3390/s20226420},
  number = {22},
  journal = {Sensors},
  publisher = {MDPI AG},
  author = {Pereira,  Felisberto and Correia,  Ricardo and Pinho,  Pedro and Lopes,  Sérgio I. and Carvalho,  Nuno Borges},
  year = {2020},
  month = nov,
  pages = {6420}
}

@article{Georgiou2018,
  title = {{The IoT Energy Challenge: A Software Perspective}},
  volume = {10},
  ISSN = {1943-0671},
  url = {http://dx.doi.org/10.1109/LES.2017.2741419},
  DOI = {10.1109/les.2017.2741419},
  number = {3},
  journal = {IEEE Embedded Systems Letters},
  publisher = {Institute of Electrical and Electronics Engineers (IEEE)},
  author = {Georgiou,  Kyriakos and Xavier-de-Souza,  Samuel and Eder,  Kerstin},
  year = {2018},
  month = sep,
  pages = {53–56}
}

@article{Bedi2018,
  title = {{Review of Internet of Things (IoT) in Electric Power and Energy Systems}},
  volume = {5},
  ISSN = {2327-4662},
  url = {http://dx.doi.org/10.1109/JIOT.2018.2802704},
  DOI = {10.1109/jiot.2018.2802704},
  number = {2},
  journal = {IEEE Internet of Things Journal},
  publisher = {Institute of Electrical and Electronics Engineers (IEEE)},
  author = {Bedi,  Guneet and Venayagamoorthy,  Ganesh Kumar and Singh,  Rajendra and Brooks,  Richard R. and Wang,  Kuang-Ching},
  year = {2018},
  month = apr,
  pages = {847–870}
}

@article{feng2020high,
  title={{High-Performance Magnesium-Carbon Nanofiber Hygroelectric Generator Based on Interface-Mediation-Enhanced Capacitive Discharging Effect}},
  author={Feng, J. and Xiao, M. and Hui, Z. and Shen, D. and Tian, Y. and Hang, C. and Duley, W. W. and Zhou, N. Y.},
  journal={ACS Applied Materials \& Interfaces},
  volume={12},
  number={21},
  pages={24289--24297},
  year={2020},
  doi={10.1021/acsami.0c04895},
  PMID={32364363}
}

@inproceedings{Lauridsen2018,
  title = {{An Empirical NB-IoT Power Consumption Model for Battery Lifetime Estimation}},
  url = {http://dx.doi.org/10.1109/VTCSpring.2018.8417653},
  DOI = {10.1109/vtcspring.2018.8417653},
  booktitle = {2018 IEEE 87th Vehicular Technology Conference (VTC Spring)},
  publisher = {IEEE},
  author = {Lauridsen,  Mads and Krigslund,  Rasmus and Rohr,  Marek and Madueno,  German},
  year = {2018},
  month = jun,
  pages = {1–5}
}

@misc{ME2108_datasheet,
  author       = {{Micro One Electronics Inc.}},
  title        = {{ME2108 DC-DC Boost Converter Datasheet}},
  year         = {2021},
  howpublished = {\url{https://www.chipsourcetek.com/DataSheet/ME2108.pdf}},
  note         = {Accessed: 2025-03-29}
}

@misc{tlv431_datasheet,
  author       = {{Diodes Incorporated}},
  title        = {{TLV431: Programmable Precision Shunt Regulator Datasheet}},
  howpublished = {\url{https://www.diodes.com/assets/Datasheets/TLV431.pdf}},
  year         = {2023},
  note         = {Accessed: 2025-03-29}
}

@misc{zxm62p02e6_datasheet,
  author       = {{Diodes Incorporated}},
  title        = {{ZXM62P02E6: 20V P-Channel Enhancement Mode MOSFET Datasheet}},
  howpublished = {\url{https://www.diodes.com/assets/Datasheets/ZXM62P02E6.pdf}},
  year         = {2023},
  note         = {Accessed: 2025-03-29}
}

@article{Sudevalayam2011,
  title = {{Energy Harvesting Sensor Nodes: Survey and Implications}},
  volume = {13},
  ISSN = {1553-877X},
  url = {http://dx.doi.org/10.1109/SURV.2011.060710.00094},
  DOI = {10.1109/surv.2011.060710.00094},
  number = {3},
  journal = {IEEE Communications Surveys \& Tutorials},
  publisher = {Institute of Electrical and Electronics Engineers (IEEE)},
  author = {Sudevalayam,  Sujesha and Kulkarni,  Purushottam},
  year = {2011},
  pages = {443–461}
}

@article{Ban2013,
  title = {{Charging and discharging electrochemical supercapacitors in the presence of both parallel leakage process and electrochemical decomposition of solvent}},
  volume = {90},
  ISSN = {0013-4686},
  url = {http://dx.doi.org/10.1016/j.electacta.2012.12.056},
  DOI = {10.1016/j.electacta.2012.12.056},
  journal = {Electrochimica Acta},
  publisher = {Elsevier BV},
  author = {Ban,  Shuai and Zhang,  Jiujun and Zhang,  Lei and Tsay,  Ken and Song,  Datong and Zou,  Xinfu},
  year = {2013},
  month = feb,
  pages = {542–549}
}

@article{Dar2024,
  title = {{Advancements in Supercapacitor electrodes and perspectives for future energy storage technologies}},
  volume = {70},
  ISSN = {0360-3199},
  url = {http://dx.doi.org/10.1016/j.ijhydene.2024.05.191},
  DOI = {10.1016/j.ijhydene.2024.05.191},
  journal = {International Journal of Hydrogen Energy},
  publisher = {Elsevier BV},
  author = {Dar,  Mohd Arif and Majid,  S.R. and Satgunam,  M and Siva,  C. and Ansari,  Shaheer and Arularasan,  P. and Rafi Ahamed,  S.},
  year = {2024},
  month = jun,
  pages = {10–28}
}

@misc{epa_fix_a_leak,
  author       = {U.S. Environmental Protection Agency},
  title        = {{Leak Facts | WaterSense}},
  year         = {2017},
  url          = {https://19january2017snapshot.epa.gov/www3/watersense/pubs/fixleak.html},
  note         = {Accessed: 2025-03-30}
}

@inproceedings{Kadu2015,
  title = {{Infrastructure Leakage Index and Challenges in Water Loss Management in Developing Countries}},
  url = {http://dx.doi.org/10.1061/9780784479162.130},
  DOI = {10.1061/9780784479162.130},
  booktitle = {World Environmental and Water Resources Congress 2015},
  publisher = {American Society of Civil Engineers},
  author = {Kadu,  Mahendra S. and Dighade,  Rajendra R.},
  year = {2015},
  month = may,
  pages = {1322–1331}
}

@article{Ali2022,
  title = {{A solution for water management and leakage detection problems using IoTs based approach}},
  volume = {18},
  ISSN = {2542-6605},
  url = {http://dx.doi.org/10.1016/j.iot.2022.100504},
  DOI = {10.1016/j.iot.2022.100504},
  journal = {Internet of Things},
  publisher = {Elsevier BV},
  author = {Ali,  Ahmed S. and Abdelmoez,  Mahmoud N. and Heshmat,  M. and Ibrahim,  Khalil},
  year = {2022},
  month = may,
  pages = {100504}
}

@article{MaheshKumar2022,
  title = {{Water Pipeline Leakage Detection and Monitoring System Using Smart Sensor with IoT}},
  volume = {2267},
  ISSN = {1742-6596},
  url = {http://dx.doi.org/10.1088/1742-6596/2267/1/012122},
  DOI = {10.1088/1742-6596/2267/1/012122},
  number = {1},
  journal = {Journal of Physics: Conference Series},
  publisher = {IOP Publishing},
  author = {Mahesh Kumar,  D and Jagadeep,  T},
  year = {2022},
  month = may,
  pages = {012122}
}

@article{Subasinghage2024,
  title = {{Supercapacitor-Assisted Energy Harvesting Systems}},
  volume = {17},
  ISSN = {1996-1073},
  url = {http://dx.doi.org/10.3390/en17153853},
  DOI = {10.3390/en17153853},
  number = {15},
  journal = {Energies},
  publisher = {MDPI AG},
  author = {Subasinghage,  Kasun and Gunawardane,  Kosala},
  year = {2024},
  month = aug,
  pages = {3853}
}

@article{Zhong2017,
  title = {{Integration of Energy Harvesting and Electrochemical Storage Devices}},
  volume = {2},
  ISSN = {2365-709X},
  url = {http://dx.doi.org/10.1002/admt.201700182},
  DOI = {10.1002/admt.201700182},
  number = {12},
  journal = {Advanced Materials Technologies},
  publisher = {Wiley},
  author = {Zhong,  Yu and Xia,  Xinhui and Mai,  Wenjie and Tu,  Jiangping and Fan,  Hong Jin},
  year = {2017},
  month = oct 
}

@manual{keysight34410A34411A,
  title        = {34410A and 34411A Multimeters},
  author       = {Keysight Technologies},
  organization = {Keysight Technologies},
  year         = {2022},
  month        = {July},
  url          = {https://www.keysight.com/ca/en/assets/7018-01326/data-sheets/5989-3738.pdf}
}

@article{Zhang2018,
  title = {{A Study on the Open Circuit Voltage and State of Charge Characterization of High Capacity Lithium-Ion Battery Under Different Temperature}},
  volume = {11},
  ISSN = {1996-1073},
  url = {http://dx.doi.org/10.3390/en11092408},
  DOI = {10.3390/en11092408},
  number = {9},
  journal = {Energies},
  publisher = {MDPI AG},
  author = {Zhang,  Ruifeng and Xia,  Bizhong and Li,  Baohua and Cao,  Libo and Lai,  Yongzhi and Zheng,  Weiwei and Wang,  Huawen and Wang,  Wei and Wang,  Mingwang},
  year = {2018},
  month = sep,
  pages = {2408}
}

@online{hawle2024,
  title = {Non-Revenue Water – Definition and Global Impact},
  author = {{Hawle}},
  year = {2024},
  url = {https://www.hawle.com/en/hawle-knowledge/basics/non-revenue-water}
}

@online{veolia2024,
  title = {Detecting Water Leaks to Reduce Potable Water Loss},
  author = {{Veolia}},
  year = {2024},
  url = {https://www.veolia.com/en/planetlive/detecting-water-leaks-reduce-potable-water-loss}
}

@techreport{un_sdg_report_2025,
  author       = {{United Nations Department of Economic and Social Affairs}},
  title        = {The Sustainable Development Goals Report 2025},
  institution  = {United Nations},
  year         = {2025},
  url          = {https://unstats.un.org/sdgs/report/2025/},
  note         = {Accessed: 2025-09-20}
}

@misc{3gpp_rel17,
  author       = {{3rd Generation Partnership Project (3GPP)}},
  title        = {{Release 17}},
  year         = {2022},
  url          = {https://www.3gpp.org/specifications-technologies/releases/release-17},
  note         = {Accessed: 2025-03-30}
}

\newpage

\vfill

\end{document}